\documentclass[useAMS,usenatbib,usegraphicx]{mn2e}

\def\simless{\mathbin{\lower 3pt\hbox
   {$\rlap{\raise 5pt\hbox{$\char'074$}}\mathchar"7218$}}} 
\def\simgreat{\mathbin{\lower 3pt\hbox
   {$\rlap{\raise 5pt\hbox{$\char'076$}}\mathchar"7218$}}} 
\def\sqdeg{~$\rm deg^2$}
\def\fluxlima{$5 \times 10^{-15} \, \rm erg \, s^{-1} \, cm^{-2}$}
\def\fluxlimb{$8 \times 10^{-15} \, \rm erg \, s^{-1} \, cm^{-2}$}
\def\arcs{\hbox{$^{\prime\prime}$}}

\title[XMM Large-Scale Structure Survey]{The XMM Large-Scale Structure
Survey: An initial sample of galaxy groups and clusters to a redshift
$z<0.6$ \thanks{Based upon observations performed at Paranal
(70.A-0283), Las Campanas and CTIO observatories and on observations
obtained with XMM-Newton, an ESA science mission with instruments and
contributions directly funded by ESA Member States and NASA}}

\author[J.P. Willis et al.]{J.P. Willis,$^{1}$\thanks{Email:
jwillis@uvic.ca} F. Pacaud,$^{2}$ I. Valtchanov,$^{2,3}$ 
M. Pierre,$^{2}$ T. Ponman,$^{4}$ A. Read,$^{5}$ \newauthor
S. Andreon,$^{6}$ B. Altieri,$^{7}$ H. Quintana,$^{8}$ S. Dos
Santos,$^{2}$ M. Birkinshaw,$^{9}$ M. Bremer,$^{9}$ \newauthor
P.-A. Duc,$^{2}$ G. Galaz,$^{8}$ E. Gosset,$^{10}$ L. Jones,$^{4}$ and
J. Surdej$^{10}$\\ $^{1}$Department of Physics and Astronomy,
University of Victoria, Elliot Building, 3800 Finnerty Road, Victoria,
BC, V8P 1A1 Canada.\\ $^{2}$CEA/Saclay, Service d'Astrophysique, F-91191,
Gif--sur--Yvette, France.\\ $^{3}$Astrophysics Group, Blackett Laboratory,
Imperial College of Science Technology and Medicine, London SW7 2BW,
UK.\\ $^{4}$School of Physics and Astronomy,
University of Birmingham, Edgebaston, Birmingham B15 2TT, UK.\\
$^{5}$Department of Physics and Astronomy, University of Leicester,
Leicester LE1 7RH, UK.\\ $^{6}$INAF -- Osservatorio Astronomico di
Brera, Milan, Italy.\\ $^{7}$ESA, Villafranca del Castillo, Spain.\\
$^{8}$Departamento de Astronom{\'i}a y Astrof{\'i}sica,
P.U. Cat{\'o}lica de Chile, Santiago, Chile.\\ $^{9}$Department of
Physics, University of Bristol, Tyndall Avenue, Bristol BS8 1TL, UK.\\
$^{10}$Universit{\'e} de Li{\`e}ge, All{\'e}e du 6 Ao{\^u}t, 17, B5C,
4000 Sart Tilman, Belgium.}

\begin{document}

\date{Accepted ????. Received ????; in original form ????}

\pagerange{\pageref{firstpage}--\pageref{lastpage}} \pubyear{????}

\maketitle

\label{firstpage}

\begin{abstract}

We present X--ray and optical spectroscopic observations of twelve
galaxy groups and clusters identified within the XMM Large--Scale
Structure (LSS) survey. Groups and clusters are selected as extended
X--ray sources from a 3.5{\sqdeg} XMM image mosaic above a flux limit
{\fluxlimb} in the [0.5--2] keV energy band. Deep $BVRI$ images and
multi--object spectroscopy confirm each source as a galaxy
concentration located within the redshift interval $0.29<z<0.56$. We
combine line--of--sight velocity dispersions with the X--ray
properties of each structure computed from a two--dimensional surface
brightness model and a single temperature fit to the XMM spectral
data. The resulting distribution of X--ray luminosity, temperature and
velocity dispersion indicate that the XMM--LSS survey is detecting
low--mass clusters and galaxy groups to redshifts $z < 0.6$. Confirmed
systems display little or no evidence for X--ray luminosity evolution
at a given X--ray temperature compared to lower redshift X--ray group
and cluster samples. A more complete understanding of these trends
will be possible with the compilation of a statistically complete
sample of galaxy groups and clusters anticipated within the continuing
XMM--LSS survey.

\end{abstract}


\section{Introduction}

Surveys of distant galaxy clusters map the distribution in the
universe of large amplitude density fluctuations, and so constrain key
cosmological parameters and permit secondary studies to determine how
X--ray gas and galaxy evolution proceeds as a function of environment.
Wide area X-ray surveys are well placed to compile statistically
well--defined samples of distant galaxy clusters because a) both
source confusion and the X-ray background are low compared to optical
searches, b) computation of the selection function and volume sampled
is straightforward and c) the selection of extended X-ray emitting
sources is sensitive to the signature of hot gas contained within
massive, gravitationally bound structures.

A number of systematic X-ray studies have extended both the maximum
redshift (i.e. the most luminous galaxy clusters) and the minimum
luminosity (i.e the least massive structures) to which X--ray clusters
can be identified. A comprehensive review is provided by Rosati,
Borgani {\&} Norman (2002). The principal aim of such surveys for
distant, X-ray emitting clusters is to determine their space density
evolution as a function of redshift and to constrain the combination
of the root mean square mass density fluctuations on 8$h^{-1}$ Mpc
scales, $\sigma_8$, and the overall matter density of the Universe
expressed as fraction of the closure density, $\Omega_{\rm M}$
(e.g. Borgani et al. 2001; Schuecker et al. 2001, Allen et al. 2003).
In addition to the study of global cosmological parameters, galaxy
clusters provide examples of dense cosmic environments in which it is
possible to study the evolution of the hot, X--ray emitting gas
(e.g. Ettori et al. 2004; Lumb et al. 2004) and to determine the
nature of the cluster galaxy populations and the physical processes
underlying observed trends in galaxy evolution (e.g. Yee et al. 1996;
Dressler et al. 1999; Andreon et al. 2004a).

Extending our current knowledge of low luminosity (i.e. low mass)
X--ray clusters represents an important challenge for the present
generation of X--ray surveys performed with the XMM and Chandra
facilities.  The local ($z<0.25$) X--ray Luminosity Function (XLF) for
galaxy clusters is currently determined to X--ray luminosities $L_X
\sim 10^{42} \ \rm erg \, s^{-1}$ in the [0.5,2] keV energy band
(Henry et al. 1992; Rosati et al. 1998; Ledlow et al. 1999; Boehringer
et al. 2002).  However, our understanding of such systems at redshifts
$0.25<z<0.8$ is largely restricted to luminosities $L_X \simgreat
10^{43.5} \ \rm erg \, s^{-1}$ (Henry et al. 1992; Burke et al. 1997;
Rosati et al. 1998; Vikhlinin et al. 1998; Ebeling et al. 2001).

Low luminosity ($10^{42} < L_X/{\rm erg \, s^{-1}} < 10^{43}$) X--ray
clusters correspond to low mass clusters and larger galaxy groups that
form a link between poorly defined ``field'' environments and X--ray
luminous/optically--rich clusters. If, as anticipated, X--ray clusters
occupying this luminosity range display X--ray temperatures $\rm T <
4$~keV, they are more likely to display the effects of
non--gravitational energy input into the Intra--Cluster Medium (ICM)
than hotter, more massive clusters (Ponman, Cannon \&\ Navarro
1999). Deviations of X--ray scaling relations from simple,
self--similar expectations have been studied for structures displaying
a relatively wide range of mass/temperature scales at $z<0.2$
(e.g. Sanderson et al. 2003). However, the study of X--ray emitting
structures -- selected over an extended temperature range -- at
 $z>0.2$ will provide an important insight into the evolution
of their X--ray emitting gas.  Though detailed X--ray studies of
galaxy clusters at  $0.2<z<0.6$ are in progress (e.g. Lumb et
al. 2004), few examples of X--ray emitting structures displaying
temperatures $\rm T<3$~keV are currently known at such
redshifts. Clearly, compilation of a sample of X--ray emitting galaxy
groups and clusters to  $z<0.6$ will greatly increase the
range of X-ray gas temperatures over which evolutionary effects can be
studied.

Low mass clusters and groups are predicted to be sites of continuing
galaxy evolution at  $z<1$ (e.g. Kaufmann 1996; Baugh et
al. 1996). When identifying such systems, it is important to note that
extended X--ray emission arises from gravitationally bound
structures. This is an important difference when X--ray selected
cluster samples are compared to optical/NIR selected cluster samples
-- whose dynamical state can only be assessed with additional velocity
data. In addition, the X--ray properties of galaxy structures
(luminosity and temperature) constrain the gravitational mass of the
emitting structure. Extending the currently known sample of galaxy
groups and low--mass clusters at  $z>0.2$ via X--ray
observations will provide an important group/cluster sample with
consistent mass ordering. A mass ordered cluster sample will permit
several detailed studies of galaxy evolution at look back times $>3-4$
Gyr to be undertaken; e.g. morphological segregation and
merger--related effects (Heldson and Ponman 2003), Butcher-Oemler
effects (Andreon and Ettori 1999) and the evolution of colour and
luminosity functions (Andreon et al. 2004a). To date, detailed galaxy
evolution studies of moderately distant $z>0.2$, X--ray selected
clusters have been performed typically for only the most X--ray bright
(i.e. massive) systems, e.g. $L_X([0.3-3.5]\rm keV)>4\times 10^{44}$
ergs s$^{-1}$ (Yee et al. 1996; Dressler 1999). Clearly an improved
sample of systems covering an extended mass interval will permit a
detailed investigation of galaxy evolution effects as a function of
changing environment.

The X-ray Multi--Mirror (XMM) Large Scale Structure (LSS) survey
(Pierre et al. 2004) is a wide area X--ray survey with the XMM
facility with the primary aim to extend detailed studies of the X-ray
cluster correlation function, currently determined at $z<0.2$ as part
of the REFLEX survey (Schuecker et al. 2001), to a redshift of unity.
However, the XMM--LSS survey features a number of secondary aims
including determination of the cosmological mass function to faint
X--ray luminosities, the evolution of cluster galaxy populations and
the evolution of the X--ray emitting gas in clusters selected over a
range of mass scales.  The nominal point source flux limit of XMM--LSS
is {\fluxlima} in the [0.5--2] keV energy band. Refregier, Valtchanov
and Pierre (2002) demonstrate that (assuming a reasonable distribution
of cluster surface brightness profiles) the approximate flux limit for
typical extended sources is \fluxlimb\ which corresponds to a X--ray
luminosity $L_X=1.2 \times 10^{43}$ ergs s$^{-1}$ for a cluster
located at a redshift $z=0.6$ and $L_X=4.2 \times 10^{43}$ ergs
s$^{-1}$ for a cluster located at a redshift $z=1$ within the adopted
cosmological model (see below).

The above aims are predicated upon the compilation of a large,
well--defined cluster catalogue. This paper describes the first
results of the XMM--LSS survey at identifying X--ray emitting clusters
at  $z<0.6$. The first clusters identified at 
$z>0.6$ are presented in Valtchanov et al. (2004). The current paper
is organised as follows; Section 2 summarises the X--ray and optical
imaging data and the methods employed to select candidate
clusters. Section 3 describes spectroscopic observations and
reductions performed for a subset of candidate clusters. Section 4
presents the determination of cluster spectroscopic properties
(redshift and line--of--sight velocity dispersion). Section 5 presents
the determination of confirmed cluster X--ray properties (surface
brightness and temperature fitting). Section 6 presents the current
conclusions for the properties of the initial $z<0.6$ sample.
Throughout this paper a Friedmann--Robertson--Walker cosmological
model, characterised by the present--day parameters $\Omega_{\rm
M}=0.3$, $\Omega_\Lambda=0.7$ and $H_0=70$ kms$^{-1}$ Mpc$^{-1}$, is
assumed. Where used, $h$ is defined as $h=H_0/ (100$
kms$^{-1}$Mpc$^{-1}$).

\section{Identifying X--ray cluster candidates}

\subsection{X-ray data reduction and source detection}
\label{subsec_xdet}

Galaxy cluster targets presented in this paper were selected from a
mosaic of overlapping XMM pointings covering a total area of
approximately 3.5{\sqdeg}. This data set represents all XMM--LSS
pointings received by August 2002 and includes 15 A0-1 10~ks exposures
and 15 Guaranteed Time (GT) 20~ks exposures obtained as part of the
XMM Medium Deep Survey (MDS).

XMM observations were reduced employing the XMM Science Analysis
System (SAS) tasks {\tt emchain} and {\tt epchain} for the MOS and pn
detectors respectively. High background periods induced by soft--proton
flaring were excluded from the event lists and raw photon images as a
function of energy band were created. The raw images for each detector
were processed employing an iterative wavelet technique and a
Poissonian noise model with a threshold of $10^{-3}$ (equivalent to $3
\sigma$ for the Gaussian case) applied to select the significant
wavelet coefficients (Starck {\&} Pierre 1998). Each wavelet filtered
image was exposure corrected and an image mask (including deviant
pixels, detector gaps and non--exposed detector regions) was created.

Source detection was performed on the wavelet filtered X--ray images
employing the {\tt SExtractor} package \cite{bertin96}.  The
discrimination between extended (cluster) and point--like sources
(mostly AGN) was achieved employing a two--constraint test based on
the half--energy radius and the SExtractor stellarity index of the
sources.  The applied procedure is the optimum method given the XMM
PSF and the Poisson nature of the signal \citep{valtchanov01}.  The
measurement of extended source properties was performed on the EPIC/pn
images as they provide the greatest sensitivity. The EPIC/MOS images
were used to discard possible artefacts resulting from edge effects
associated with the pn CCDs.

\subsection{Optical imaging}

The selection of potential galaxy members within each candidate
cluster was performed employing moderately deep $BVRI$ images from the
CFH12k camera on the Canada France Hawaii Telescope (CFHT) obtained as
part of the VIRMOS deep imaging survey (McCracken et
al. 2003). Observations were processed employing the
Terapix\footnote{htp://terapix.iap.fr} data reduction pipeline to
produce an astrometric and photometric image data set. Object
catalogues were produced using {\tt SExtractor}. Catalogue detection
thresholds as a function of photometric band are displayed in Table
\ref{tab_vvds}.

\begin{table}
\caption{Detection thresholds as a function of photometric
passband for objects detected in the optical images.}
\label{tab_vvds}
\begin{tabular}{lc}
\hline
Filter & Detection threshold \\
& (50\% AB magnitude completeness \\
& limit for stellar sources)\\
\hline
\hline
$B$ & 26.5 \\
$V$ & 26.0 \\
$R$ & 26.0 \\
$I$ & 25.4 \\
\hline
\end{tabular}
\end{table}

\subsection{Selecting candidate clusters and member galaxies}
\label{sec_selcandclus}

The identification of galaxy clusters over an extended redshift
interval in X--ray images is limited by the ability of the XMM
facility to identify extended cluster X--ray emission in a 10ks
exposure.  The Half--Energy Width (HEW) of a on--axis point source is
approximately 15\arcsec\ at 1.5 keV. However, the HEW displays marked
local variations resulting from off--axis angle, vignetting and
detector gaps.  The on--axis HEW corresponds to a projected transverse
distance of 120 kpc at a redshift $z=1$ within the assumed
cosmological model. Although this HEW is sufficient to resolve the
extended emission from massive galaxy clusters to $z \simgreat 1$, the
effect of low central surface brightness, leading to a truncated
detectable cluster extension, can lead to a cluster being erroneously
identified as an unresolved object. Therefore both distant clusters
and intrinsically compact clusters at all redshifts may potentially
appear as only marginally resolved or unresolved sources in XMM--LSS
X--ray mosaics. A quantitative assessment of the cluster X--ray
selection function will form the subject of a future paper (Pacaud et
al. 2005).

The XMM--LSS incorporates a number of different solutions to the
problem of cluster identification, e.g. correlation of extended X--ray
sources with optical galaxy structures (this paper and Valtchanov et
al. 2004), investigation of the X--ray properties of optically
identified structures (and vice versa; Andreon et al. 2004b) and the
investigation of extended X--ray sources lacking optical counterparts
together with unresolved X-ray sources associated with faint optical
structures (Andreon et al. 2004b).

Analysis of the first $3.5 \ \rm deg^2$ of the XMM--LSS survey led to
the identification of 55 extended sources with fluxes greater than
\fluxlimb\ which are extended according to the criteria detailed in
Section \ref{subsec_xdet}. The optical imaging data corresponding to a
$7\arcmin \times 7\arcmin$ field\footnote{A practical limit determined
by the field of view of the multi--object spectroscopic facilities
employed to observe cluster candidates (see Section
\ref{sec_specobs}). The field size is large compared to the extent of
the X--ray emission and obviates any requirement to adjust the field
centre to maximise the number of candidate cluster members.}  centred
upon the location of each extended X--ray source was analysed for the
presence of a significant galaxy structure showing a well defined, red
colour sequence (Stanford et al. 1998; Kodama et al. 1998).  Candidate
cluster members were selected by inspection of the available $BVRI$
photometry of objects identified within the field of each extended
X--ray source. Colour magnitude thresholds were applied interactively
in order to enhance galaxy structures when viewed in a $VRI$
pseudo--colour image of each field with X--ray contours superposed
from the wavelet--filtered X--ray image. One exception to this
procedure was cluster candidate XLSS J022722.3-032141 (see Table
\ref{tab_clsobs}); the optical data for this candidate cluster
consisted of the $I$--band VLT/FORS2 pre-image obtained to define slit
locations. Candidate cluster members were selected to include all
galaxies up to 1.5 magnitudes fainter than the bright ($I=17.1$),
central galaxy associated with the extended X--ray source.

The sample of candidate cluster members generated by the above
procedures was used to design spectroscopic masks for each candidate
cluster field. Two multi--slit masks were created for each candidate
cluster with brighter galaxies given higher priority in the slit
assignment procedure. Unused regions of each multi--slit mask were
employed to sample the population of unresolved X--ray sources with
bright ($R<23$) optical counterparts. Further discussion of this
additional sample will appear elsewhere.

\section{Spectroscopic observations}
\label{sec_specobs}

The candidate cluster sample was observed by the Las Campanas
Observatory Baade telescope with the Low Dispersion Survey
Spectrograph (LDSS2) during 4-5 October 2002 and the European Southern
Observatory Very Large Telescope (VLT) with Focal Reducing
Spectrograph (FORS2) during 9-12 October 2002. Each instrument (LDSS2
and FORS2) is a focal reducing spectrograph with both an imaging and a
multi-object spectroscopy (MOS) capability. In each case, MOS
observations are performed using slit masks mounted in the instrument
focal plane. Details of which cluster candidate was observed with
which telescope plus instrument configuration are provided in Table
\ref{tab_clsobs}. The effective wavelength interval, pixel sampling
and spectral resolution generated by each instrument combination are
indicated in Table \ref{tab_iperf}.

\begin{table*}
%
\begin{minipage}{180mm}
\caption{Observing log of candidate groups and clusters.}
\label{tab_clsobs}
\begin{tabular}{@{}clcccccc}
\hline
ID\footnote{In the following text, all clusters are referred to via
the reference XLSSC plus the identification number, e.g. XLSSC 006,
etc. Clusters 001--005 correspond to redshift $z>0.6$ clusters
presented in Valtchanov et al. (2004).} & Cluster & Right
Ascension\footnote{Positions are J2000.0.} & Declination & Instrument
& $\rm Grism + Filter$ & {\#} of masks & Total exposure time \\
&&&&&&& per mask (seconds) \\
\hline
\hline
006 & XLSSUJ022145.2-034614 & 02:21:45.22 & $-03$:46:14.1 & FORS2 & $\rm 300V + GG435$ & 2 & $2700 + 2400$ \\
007 & XLSSUJ022406.0-035511 & 02:24:05.95 & $-03$:55:11.4 & FORS2 & $\rm 600RI+ GG435$ & 2 & $2400 + 2400$ \\
008 & XLSSUJ022520.7-034800 & 02:25:20.71 & $-03$:48:00.0 & FORS2 & $\rm 600RI+ GG435$ & 2 & $1200 + 1200$ \\
009 & XLSSUJ022644.2-034042 & 02:26:44.21 & $-03$:40:41.8 & FORS2 & $\rm 300V + GG435$ & 2 & $1800 + 1800$ \\
010 & XLSSUJ022722.2-032137 & 02:27:22.16 & $-03$:21:37.0 & FORS2 & $\rm 600RI+ GG435$ & 1 & $600$         \\
012 & XLSSUJ022827.5-042554 & 02:28:27.47 & $-04$:25:54.3 & LDSS2         & medium-red & 2 & $1800 + 1800$ \\
013 & XLSSUJ022726.0-043213 & 02:27:25.98 & $-04$:32:13.1 & LDSS2         & medium-red & 2 & $900 + 900$   \\
014 & XLSSUJ022633.9-040348 & 02:26:33.87 & $-04$:03:48.0 & LDSS2         & medium-red & 2 & $1800 + 1800$ \\
016 & XLSSUJ022829.0-045932 & 02:28:29.03 & $-04$:59:32.2 & LDSS2         & medium-red & 2 & $600 + 900$   \\
017 & XLSSUJ022628.2-045948 & 02:26:28.19 & $-04$:59:48.1 & LDSS2         & medium-red & 2 & $1800 + 1800$ \\
018 & XLSSUJ022401.5-050525 & 02:24:01.46 & $-05$:05:24.8 & LDSS2         & medium-red & 2 & $2400 + 2400$ \\
020 & XLSSUJ022627.0-050008 & 02:26:27.08 & $-05$:00:08.4 & LDSS2         & medium-red & 2 & $1800 + 1800$ \\
\hline
\end{tabular}
\end{minipage}
\end{table*}

\begin{table*}
\begin{minipage}{180mm}
\caption{Instrumental characteristics for each spectrograph
configuration employed during the observations}.
\label{tab_iperf}
%
%
\begin{tabular}{lcccc}
\hline 
Instrument & $\rm Grism + Filter$ & Wavelength &
Pixel sampling  & Spectral \\
&& interval (\AA) & ($\rm \AA \, {pix}^{-1}$) &
resolution\footnote{Estimated for each spectrograph via the mean
full--width at half--maximum of the HeI5876 arc emission line.All
spectral observations were performed with a slit width of
approximately 1\farcs4.} (\AA) \\
\hline
\hline
FORS2 & $\rm 300V + GG435$ & 4000--9000 & 3.2 & 14 \\
FORS2 & $\rm 600RI+ GG435$ & 5000--8500 & 1.6 & 7  \\
LDSS2 & medium--red        & 4000--9000 & 5.1 & 14 \\
\hline
\end{tabular}
\end{minipage}
\end{table*}

Spectroscopic observations were reduced employing standard data
reduction procedures within {\tt IRAF}\footnote{IRAF is distributed by
the National Optical Astronomy Observatories, which are operated by
the Association of Universities for Research in Astronomy, Inc., under
cooperative agreement with the National Science Foundation.}: a zero
level, flat--field and cosmic ray correction was applied to all MOS
observations prior to the identification, sky subtraction and
extraction of individual spectral traces employing the {\tt apextract}
package. The dispersion solution for each extracted spectrum was
determined employing HeNeAr lamp exposures and all data spectra were
resampled to a linear wavelength scale. A single spectrophotometric
standard star from the atlas of Hamuy et al. (1992, 1994) was observed
during each night and was employed to correct for the relative
instrumental efficiency as a function of wavelength. Removal of the
relative instrumental efficiency as a function of wavelength does not
affect the later determination of galaxy redshifts via
cross--correlation analysis. However, it does permit spectra to be
displayed on a relative spectral flux scale that aids the visual
assessment of low quality spectra.

\subsection{Spectral classification and redshift determination}

In order to confirm the redshift of each candidate cluster, the
spectroscopic sample generated for each field was constructed to
maximise the number of potential cluster members according to the
photometric criteria described in Section
\ref{sec_selcandclus}. Though these criteria were constructed in order
to identify the characteristic colour signature generated by an
overdensity of early--type galaxies at a particular redshift, it is
probable that the spectral sample generated for each candidate cluster
is contaminated by the presence of galaxies within the target field
that are (gravitationally) unassociated with the cluster and with
stars misidentified as galaxies. In order to address this issue and to
classify each candidate cluster member, all reduced spectra were
inspected visually to identify contaminating stars and to provide an
initial estimate of galaxy redshifts based upon the identification of
prominent features. Individual spectra were then cross--correlated
with a representative early--type galaxy template (Kinney et al. 1996)
employing the {\tt IRAF} routine {\tt xcsao} (Tonry {\&} Davis
1979). The cross--correlation procedure was performed interactively in
order to improve the identification of a reliable cross--correlation
peak. Spectral regions corresponding to the observed locations of
prominent night sky emission features and regions of strong
atmospheric absorption were masked within the cross--correlation
analysis. Computed redshift values have not been corrected to a
heliocentric velocity scale.

Errors in the cross--correlation velocity returned by {\tt xcsao} are
computed based upon the fitted peak height and the antisymmetric noise
component associated with the identified cross--correlation peak
(Tonry {\&} Davis 1979; Heavens 1993). The typical median velocity
error computed for spectra observed with each spectrograph combination
described in Table \ref{tab_iperf} is 75 kms$^{-1}$ for FORS2$+$600RI
and 150 kms$^{-1}$ for FORS2$+$300V and LDSS2/medium--red.  The random
error associated with uncertainties in the dispersion solution applied
to each spectrum was characterised via determination of the error in
the wavelength location of prominent emission features in the night
sky spectrum associated with each data spectrum when compared to their
reference values. The distribution of wavelength residuals are
considered in velocity space and the {\tt rostat} statistics package
(Beers, Flynn {\&} Gebhardt 1990; see Section \ref{sec_clsprop}) is
employed to calculate the bi-weight mean and scale for each cluster
field. Typical values (for all spectrograph settings) of the mean
wavelength shift and dispersion computed via this method are $<\pm100$
kms$^{-1}$ and $<50$ kms$^{-1}$ respectively. Radial velocities (from
which redshift of each galaxy is determined) are corrected for the
mean velocity residual in each field and we further assume that the
distribution of errors in the cross--correlation velocities and in the
dispersion solutions to be Gaussian and therefore calculate a total
uncertainty in the corrected radial velocity by combining these two
sources of error in quadrature.

\section{Determination of cluster spectroscopic properties}
\label{sec_clsprop}

The nature of each candidate cluster contained within the
spectroscopic sample is assessed employing the available X--ray and
optical images and the spectroscopic information accumulated for each
field. The field of each candidate cluster is inspected visually
employing a composite image containing the CFH12k $R$--band greyscale
image, X--ray contours derived from the wavelet--filtered XMM mosaic
and the available redshifts of all galaxies contained within the field
(see Figure \ref{all_clus_plot}).  The redshift distribution generated
by all spectroscopic redshifts obtained in each candidate cluster
field is also displayed. To illustrate redshift space overdensities,
the redshift density computed by applying an adaptive kernel
(Silverman 1986) to the redshift data is also shown.  This process
provides an initial estimate of the cluster redshift via the
identification of three dimensional (position and redshift) structures
associated with the extended X-ray emission. This redshift estimate is
then employed to select the corresponding peak in the redshift
histogram of the field. In the case of clusters XLSSC 017 and 020, the
redshift of each extended X-ray source was determined by determining
the spatial barycentre of each redshift peak displayed in Figure 1 and
assigning the redshift grouping closest to each X-ray source as the
cluster redshift.

The typically small number ($<20$) of objects observed
spectroscopically in each cluster field limits the usefulness of any
assessment of spectroscopic completeness and redshift confirmation
frequency. However, the spectroscopic redshift reported for each
cluster is in excellent agreement ($\Delta z < 0.02$) with the
redshift determined independently from the location of the red
envelope of the corresponding cluster colour magnitude relation in
$R-z$ colour space (Andreon et al. 2004a).

\begin{figure*}
\centering
\includegraphics[width=16.5cm]{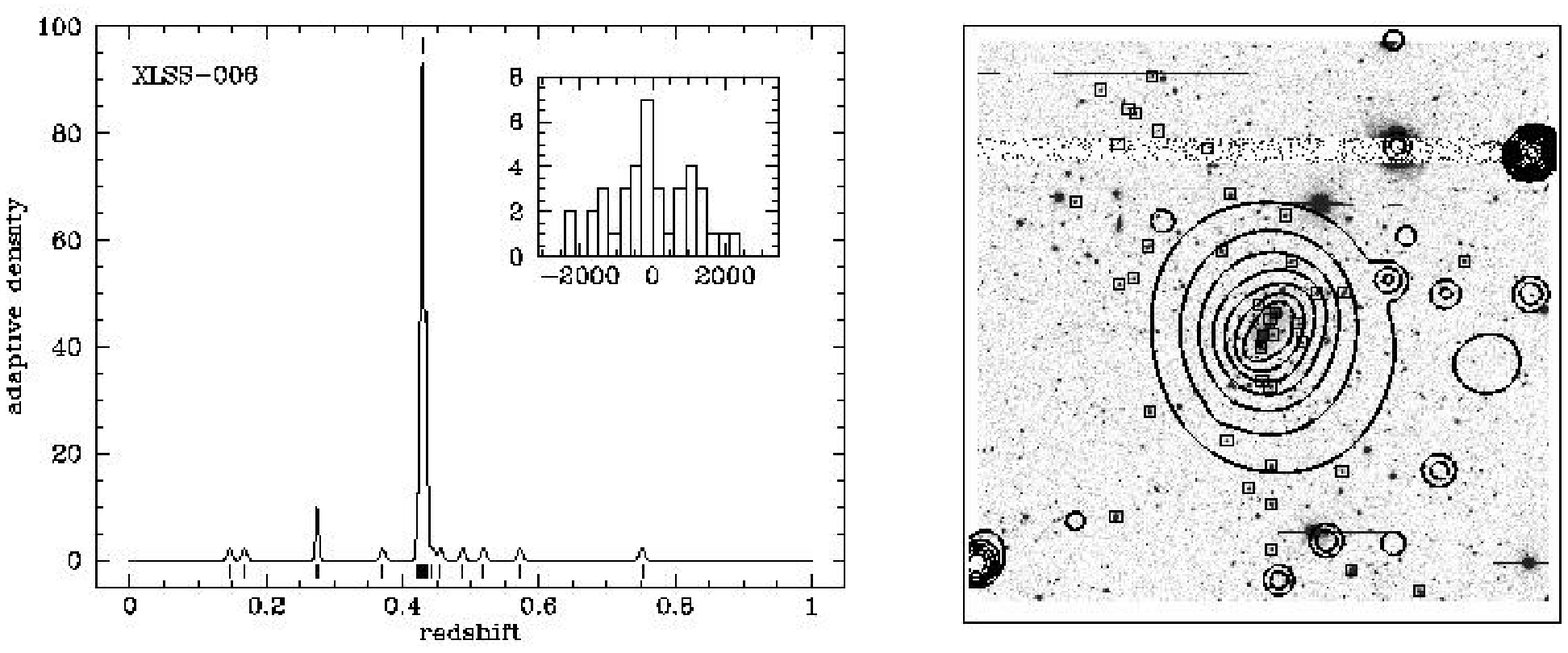}
\includegraphics[width=16.5cm]{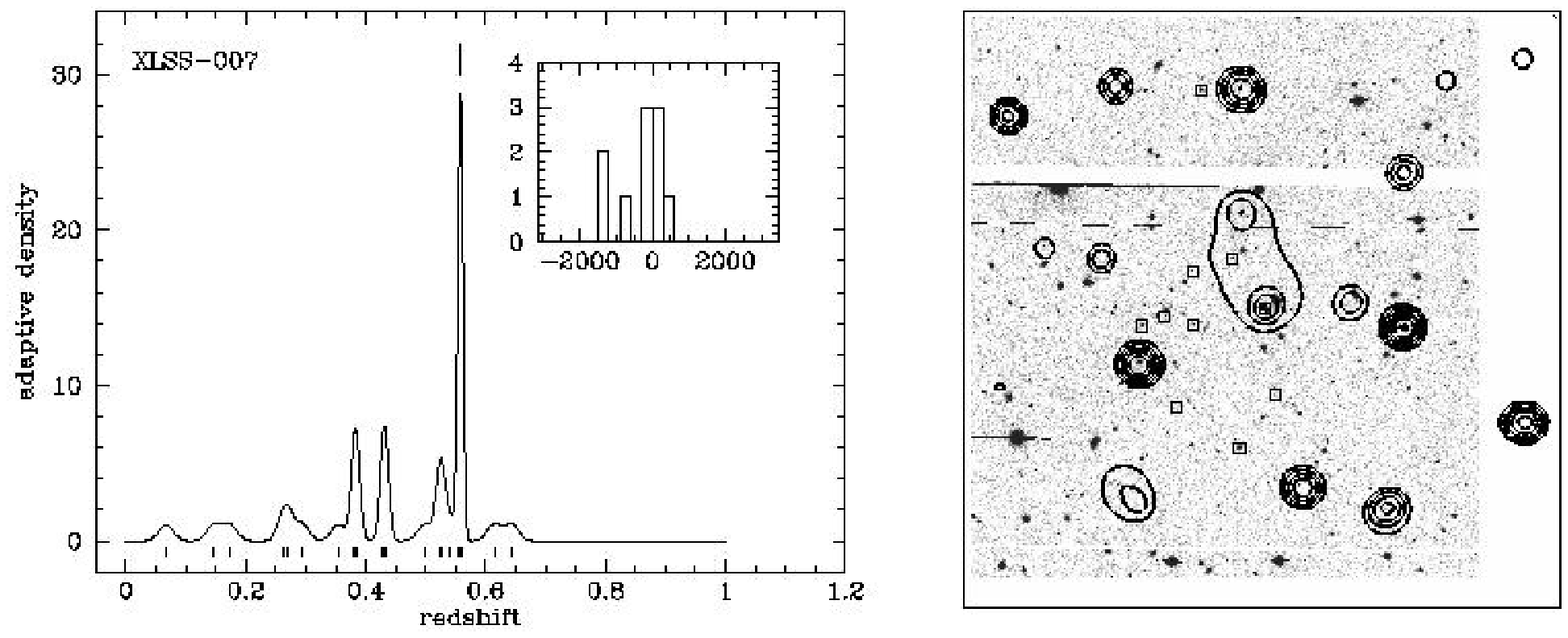}
\includegraphics[width=16.5cm]{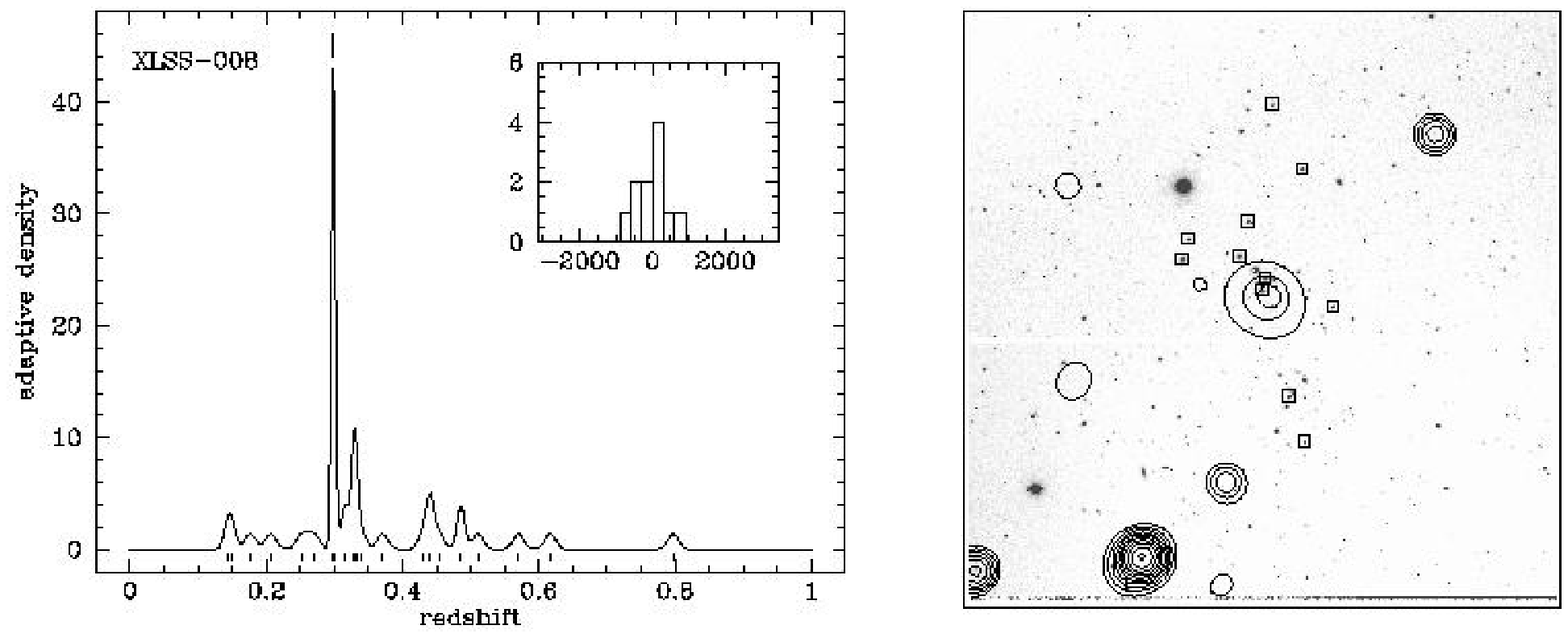}

\caption{Optical images, X-ray contours and spectroscopic information
for each cluster presented in Table \ref{tab_clsprop}. Left panel:
adaptively smoothed density profile versus redshift. Individual
redshifts are marked by short vertical lines below the density
curve. A short vertical line above the density curve marks the
redshift location of the cluster. The inset histogram displays the
rest--frame velocity distribution of confirmed cluster members. Right
panel: a $7\arcmin \times 7\arcmin$ $R$--band image centred on the
extended X-ray source. Wavelet filtered X-ray contours are overplotted
and squares indicate the position of confirmed cluster members. The
X-ray contours typically run from two times the background level in
each frame to 5 photons/pixel with 10 logarithmic levels. Note that
the optical images form a heterogeneous data set and are presented to
indicate the visual appearance of each cluster. All optical images are
orientated with North up and East left. Where more than one cluster is
detected in the same field a black arrow indicates the location of
cluster being considered.}
\label{all_clus_plot}
\end{figure*}

\setcounter{figure}{0}

\begin{figure*}
\centering
\includegraphics[width=16.5cm]{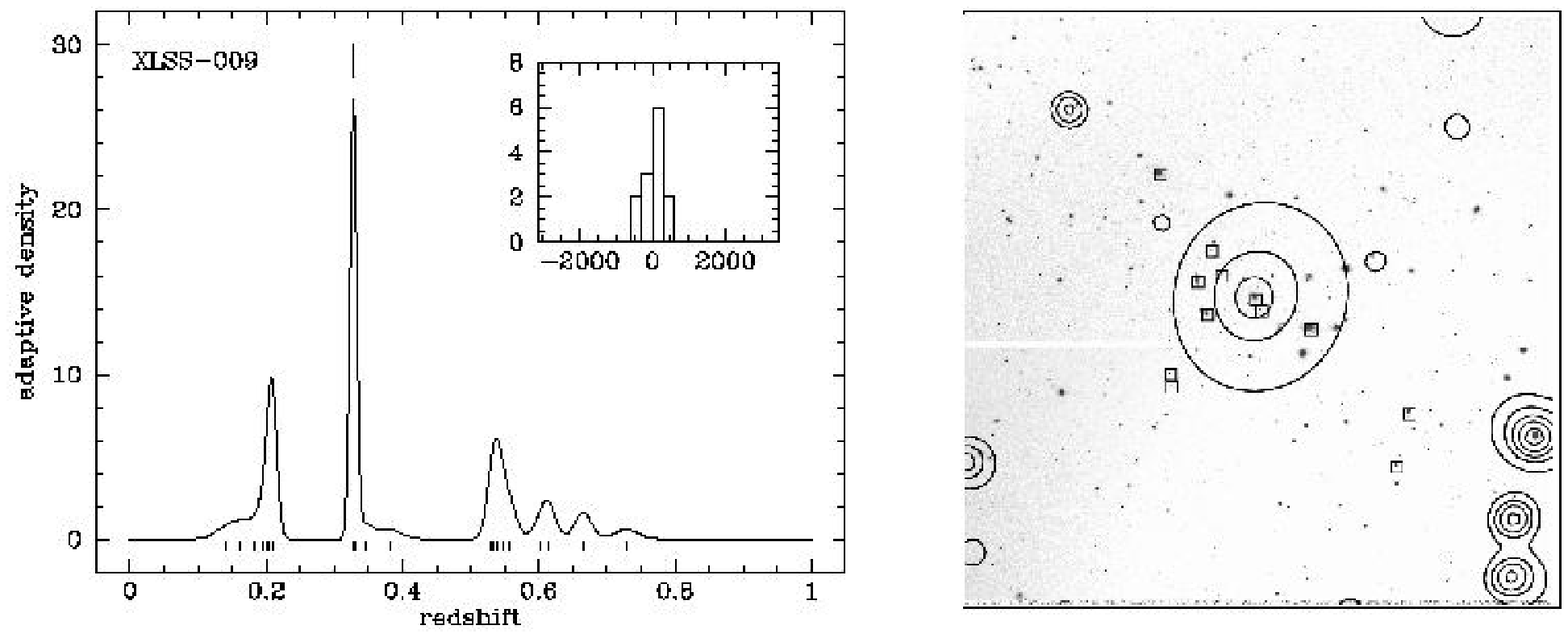}
\includegraphics[width=16.5cm]{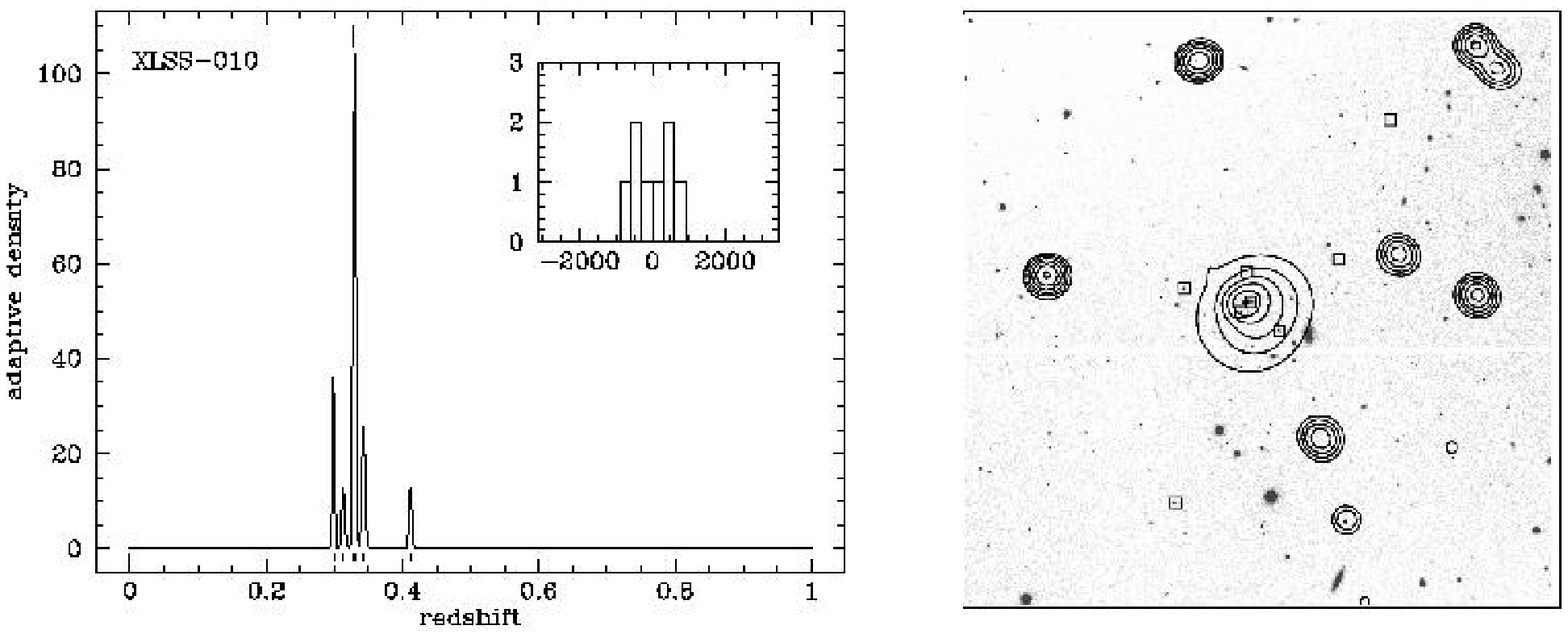}
\includegraphics[width=16.5cm]{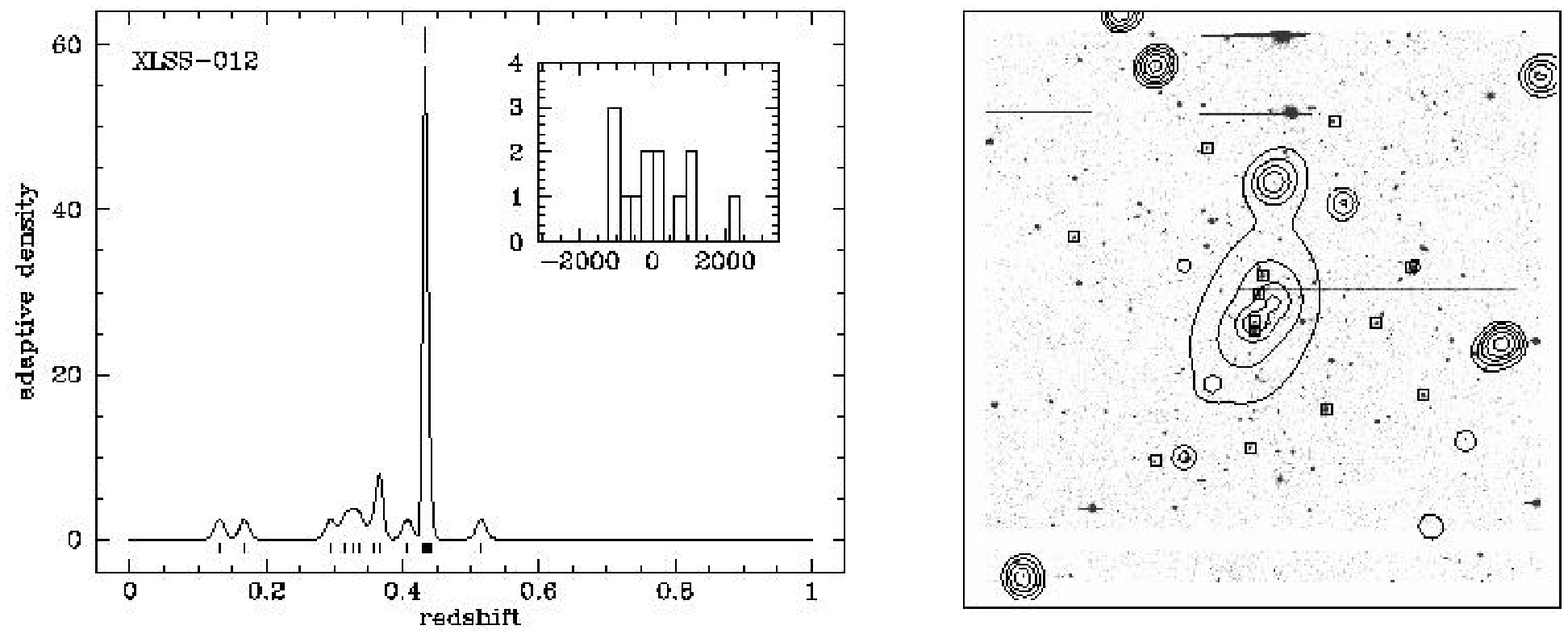}

\caption{Continued.}
\end{figure*}

\setcounter{figure}{0}

\begin{figure*}
\centering
\includegraphics[width=16.5cm]{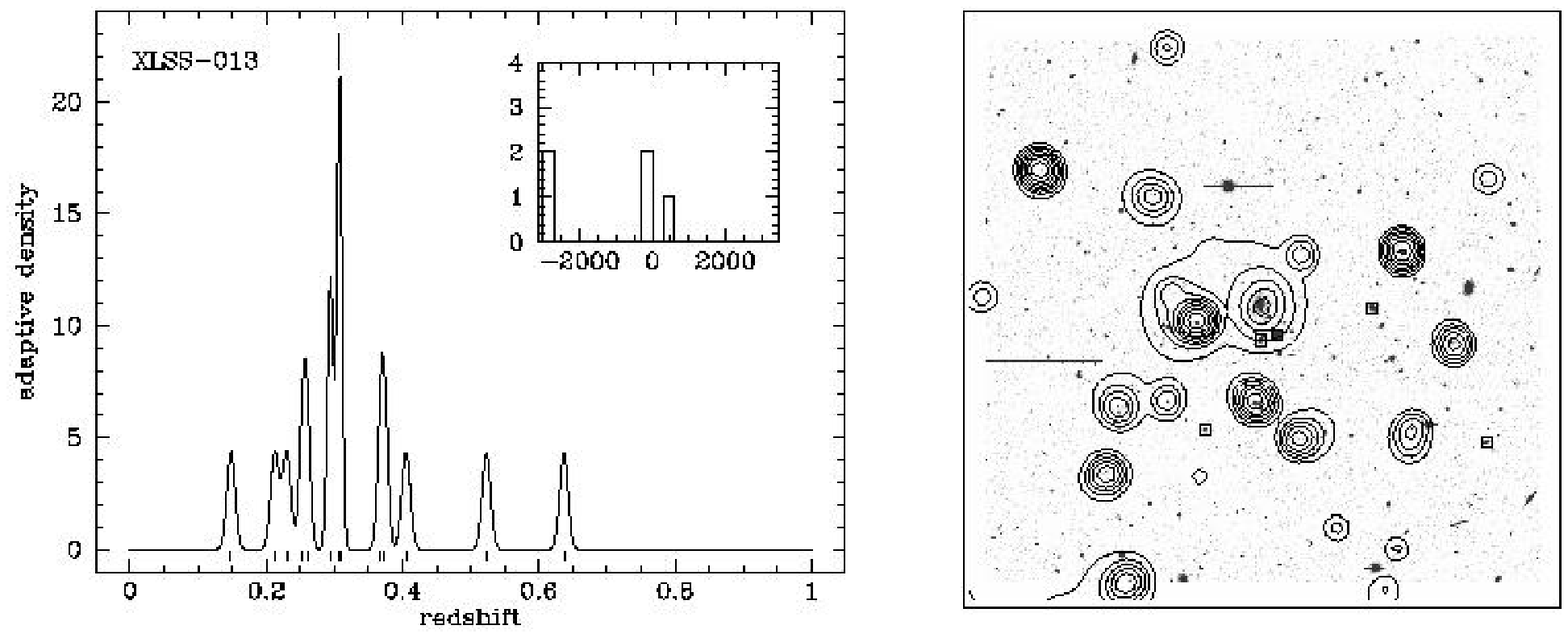}
\includegraphics[width=16.5cm]{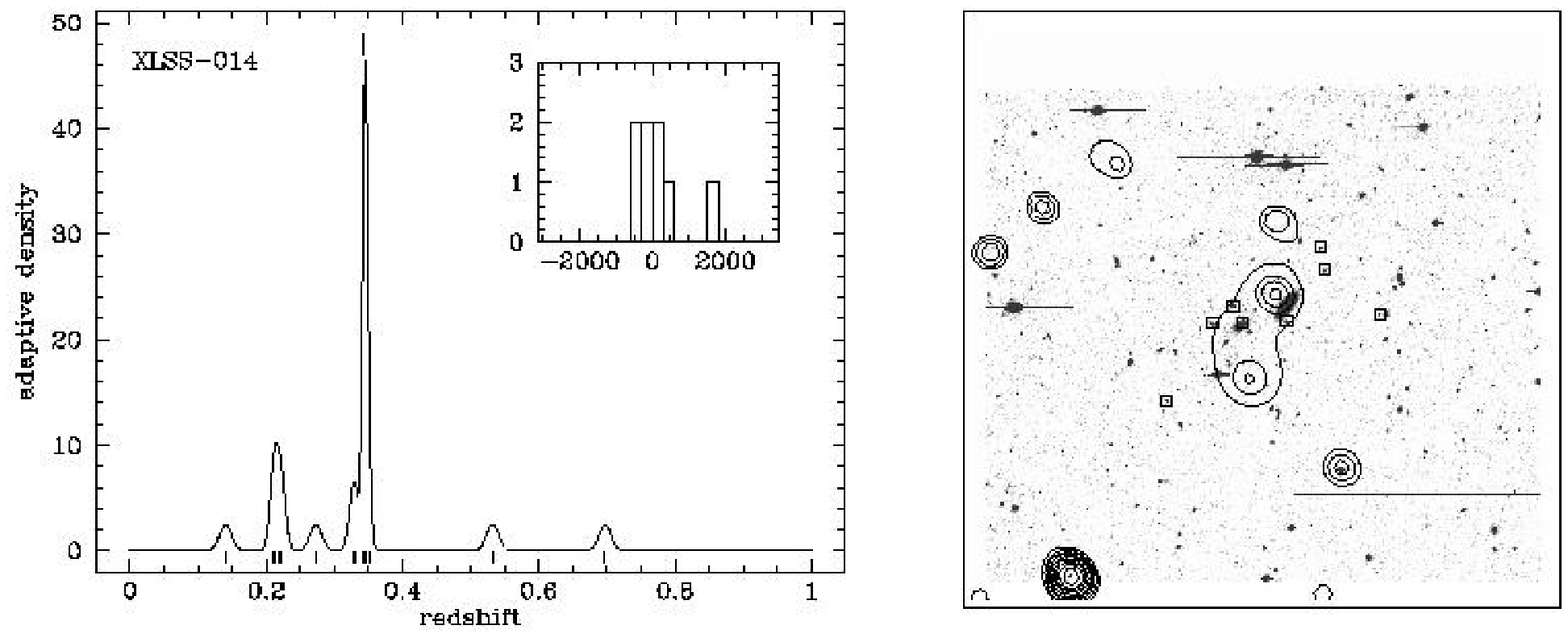}
\includegraphics[width=16.5cm]{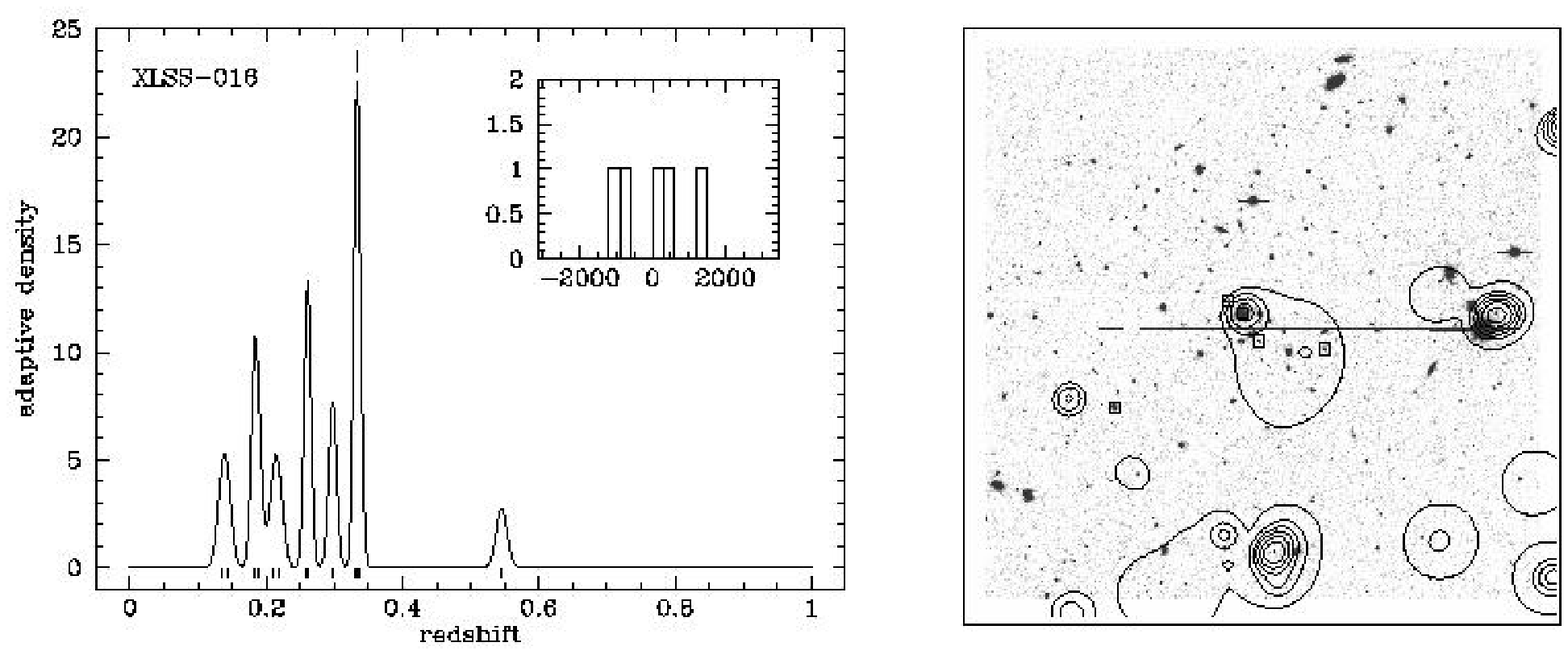}

\caption{Continued.}
\end{figure*}

\setcounter{figure}{0}

\begin{figure*}
\centering
\includegraphics[width=16.5cm]{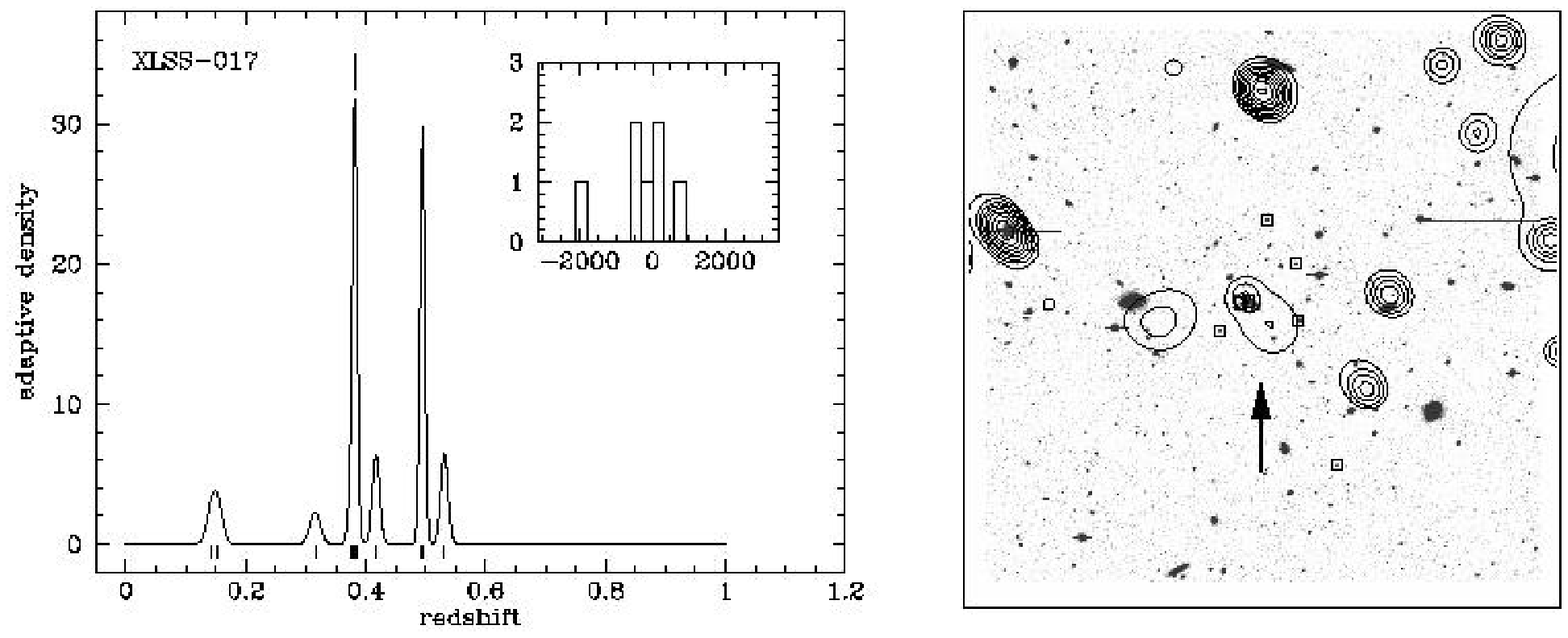}
\includegraphics[width=16.5cm]{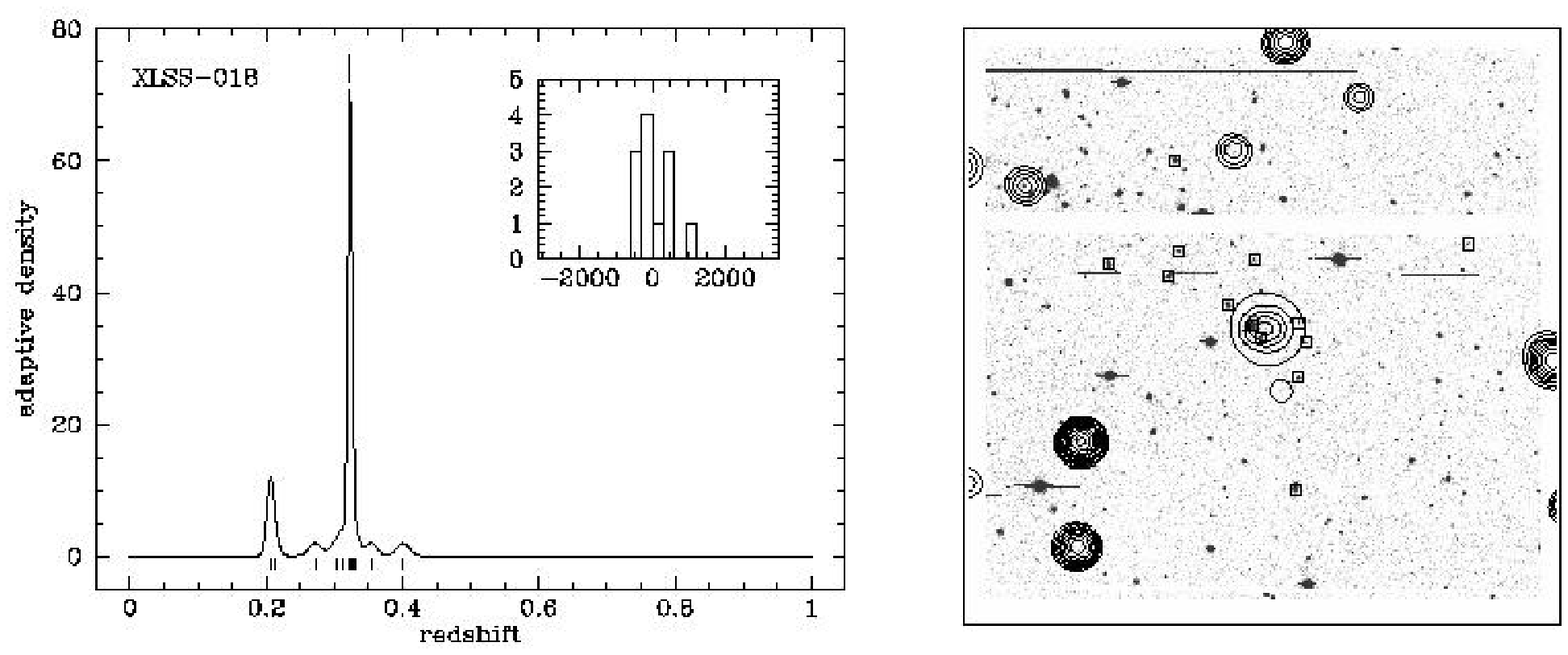}
\includegraphics[width=16.5cm]{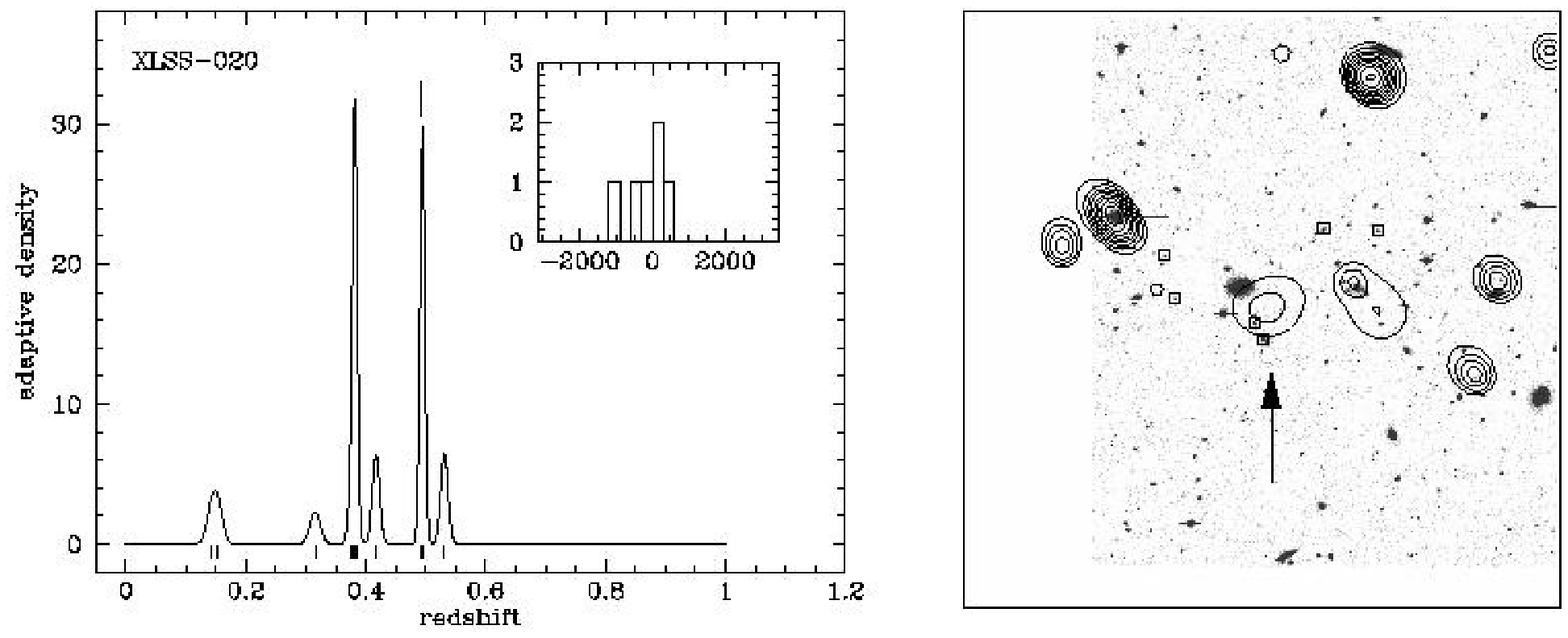}

\caption{Continued.}
\end{figure*}

The sample of cluster members is selected in radial velocity space
employing an iterative method similar to that of Lubin, Oke {\&}
Postman (2002): the initial cluster sample is selected to lie within
the redshift interval $\Delta z = \pm 0.06$ of the estimated cluster
redshift. Radial velocities relative to the cluster centre are
calculated within the cluster rest frame, i.e. $\Delta v = c \,
(z-{\bar{z}})/(1+{\bar{z}})$ where $\bar z$ is the median redshift
within the specified interval. The bi--weight mean and scale of the
radial velocity distribution within this interval is computed using
{\tt rostat} and galaxies that display a velocity difference relative
to the central location of greater than $3500 \rm kms^{-1}$ or three
scale measures, are rejected and the statistical measures
recalculated. This procedure is repeated until no further galaxies are
rejected. Errors in the bi--weight mean and scale are estimated
employing a bootstrap or jacknife calculation with 10,000 resamplings
for clusters with greater than or less than 10 confirmed members
respectively. Estimates of the bi--weighted mean radial velocity and
line--of--sight velocity dispersion of each cluster, and the
associated uncertainties, are corrected for biases arising from
measurement errors employing the prescription of Danese et al. (1980).

Although galaxy redshift and positional information is compared to the
location of the X--ray source to define the initial cluster redshift
input to the velocity search algorithm, no additional spatial
filtering of potential galaxy cluster members is performed. However,
the projected transverse distances sampled by the detector fields of
view employed to perform the spectroscopy vary between 1.5 and 2
$h^{-1}$ Mpc for clusters located over the redshift interval
$0.3<z<0.6$. Studies of velocity dispersion gradients in both local
(Girardi et al. 1996) and distant (Borgani et al. 1999) X--ray
clusters -- albeit hotter/more luminous systems than those presented
in the current paper -- indicate that integrated velocity dispersion
profiles typically converge within radii $r<1-2 \, h^{-1}$ Mpc of the
X--ray cluster centre.  Upon initial inspection, the cluster regions
sampled by the projected field of view sampled by each telescope plus
spectrograph combination would appear to be well matched to the
convergent velocity profile of typical hot/luminous clusters. However,
inspection of Figure \ref{all_clus_plot} indicates that cluster
galaxies are typically confirmed within the central regions in each
field -- a strategy required by the necessity to confirm the redshift
of galaxy structures near the X--ray source.  Though the XMM--LSS
clusters presented in this paper are typically cooler (i.e. less
massive) than those presented by Girardi et al. (1996) and Borgani et
al. (1999), and may reasonably be assumed to be intrinsically less
extensive, an unknown and potentially significant uncertainty is
associated with the assumption that the computed velocity dispersion
figures represent the {\it convergent} velocity dispersion for each
cluster. The resulting spectroscopic properties of all cluster
candidates listed in Table \ref{tab_clsobs} are given in Table
\ref{tab_clsprop}.

\begin{table}
\begin{minipage}{70mm}
\centering
\caption{Spectroscopic properties of all redshift $z < 0.6$ groups and
clusters}.
\label{tab_clsprop}
\begin{tabular}{lccl}
\hline 
Cluster & Redshift\footnote{The uncertainty associated with cluster redshifts is less than $\Delta z = 0.001$ in all cases.} & {\#} of members & $\sigma_v$\footnote{Velocity dispersion uncertainties are quoted at the 68{\%} confidence level.} \\
&&&($\rm kms^{-1}$)\\
\hline
\hline
XLSSC 006  & 0.429 &  39 &  $821_{-74}^{+92}$\\
XLSSC 007  & 0.558 &  10 &  $323_{-191}^{+178}$\\
XLSSC 008  & 0.298 &  11 &  $351_{-35}^{+98}$\\
XLSSC 009  & 0.327 &  13 &  $232_{-31}^{+60}$\\
XLSSC 010  & 0.329 &   8 &  $420 \pm 72$\\
XLSSC 012  & 0.433 &  13 &  $694_{-91}^{+204}$\\
XLSSC 013\footnote{The available data for cluster 013 does not generate a well--defined velocity dispersion.}  & 0.307 &   5 & N/A\\
XLSSC 014  & 0.344 &   8 &  $416 \pm 246$\\
XLSSC 016  & 0.332 &   5 &  $703 \pm 266$\\
XLSSC 017  & 0.382 &   7 &  $571 \pm 282$\\
XLSSC 018  & 0.322 &  12 &  $342_{-35}^{+104}$\\
XLSSC 020  & 0.494 &   6 & $ 265_{-146 }^{+240}$\\
\hline
\end{tabular}
\end{minipage}
\end{table}

\section{Determination of group and cluster X--ray properties}
\label{sec_x_prop}

In order to determine the nature of the spectroscopically confirmed
clusters presented in this paper, additional analyses of the available
XMM data were performed to characterise the spatial and spectral
properties of the X--ray emitting gas. When combined with the optical
redshift and line--of--sight velocity dispersion (where available)
information, these X--ray measures permit a comparison of the cluster
sample with lower redshift samples and in particular permit the
approximate mass interval occupied by $z<0.6$ XMM--LSS clusters to be
understood.

\subsection[]{Morphological properties}

The X--ray surface brightness distribution of each cluster was
modelled employing a circular $\beta$-model, of the form
\begin{equation}
{
f(r) = \frac{A}{[1+(r/r_{0})^{2}]^{\, \alpha}}, 
}
\end{equation}
where the coordinate $r$ is measured in arcseconds with respect to the
centre of the X-ray photon distribution, $r_{0}$ is the core radius,
$A$ is the amplitude at $r=0$, and $\alpha = 3 \, \beta -
\frac{1}{2}$.

Images and exposure maps for each cluster field were created for the
three EPIC instruments (MOS1, MOS2 and pn) separately in the
[0.5--2]~keV energy band. Images of the appropriate PSF were created
(using SAS-calview), with the appropriate energy weighting together
with the off-axis and azimuthal angles of the source. Square regions
(of sizes ranging between 175\arcs\ and 500\arcs\ on a side) were
selected around each source and around a nearby, source-free,
background region. Mask images were also created and employed to
remove from further analysis regions associated with chip gaps and
serendipitous point sources lying close to each cluster source.

We used the {\it Sherpa} package from the {\it CIAO} analysis
system to fit a model of the form given by Equation 1 to the X--ray data
for each EPIC instrument. The quality of fit to the three instruments
was optimised using the Cash statistic, providing a maximum likelihood
fit, which accounts properly for the Poissonian nature of the data.
Each model incorporates a flat background model
(where the background level is determined employing the associated
background region) and a $\beta$--model convolved with the appropriate
PSF. For each of the three instrument models determined for each
cluster, the values of the core radius $r_{0}$, position ($x_{0}$,
$y_{0}$), and slope $\alpha$ were constrained to be identical. Only
the normalisation for each model was permitted to vary.

\begin{figure*}
\centering
\includegraphics[height=10cm]{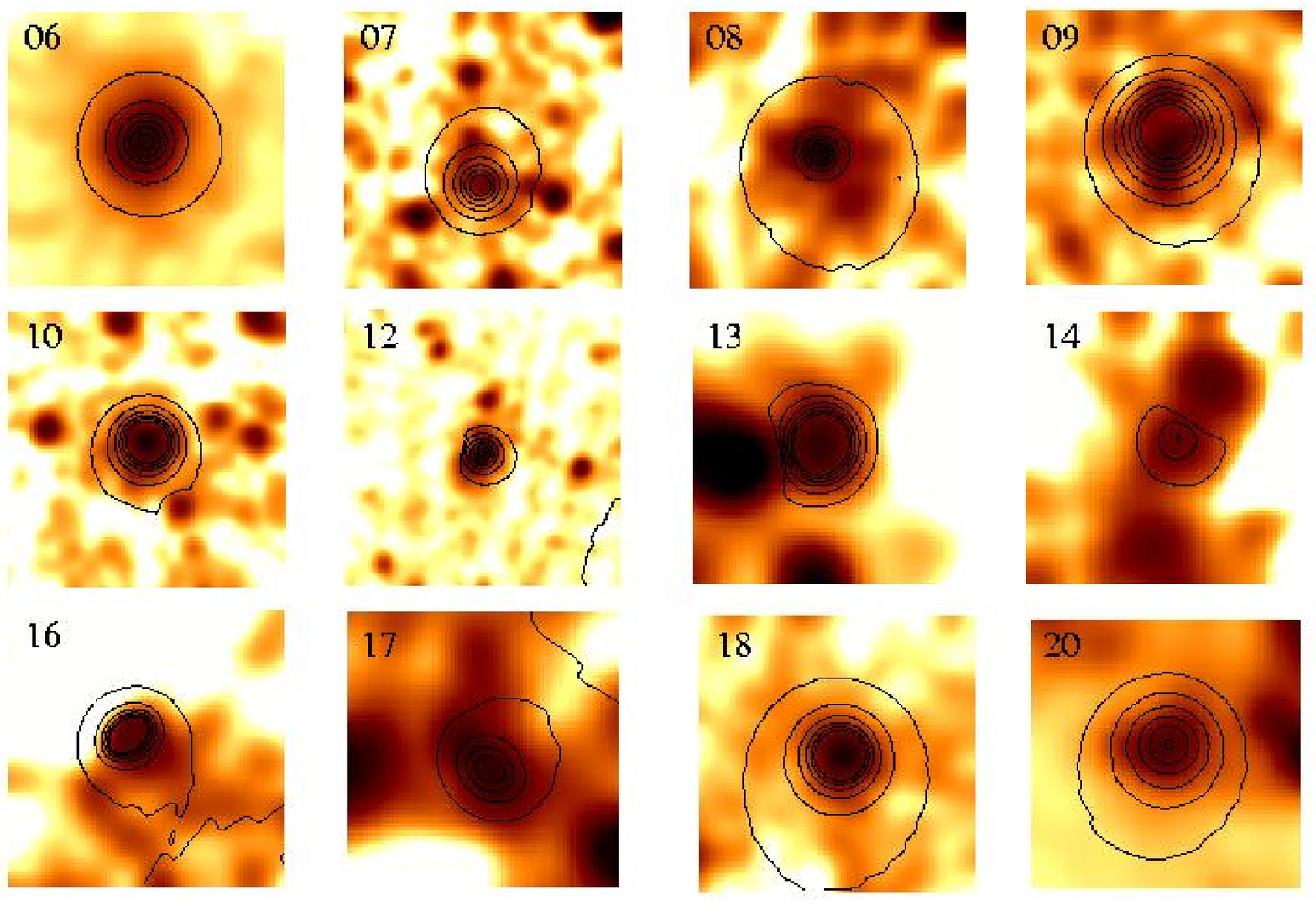}
\caption{Model surface brightness distributions for the X--ray group
and cluster sample. Contours represent the best--fitting model
convolved with the detector response and are overplotted on the full
(pn$+$MOS1$+$MOS2) Gaussian smoothed X--ray emission corresponding to
the [0.5--2] keV energy band. The Gaussian smoothing scale (i.e. the
standard sigma) is 5 pixels or 12\farcs5.The linearly spaced contours
represent the MOS1 model (only the normalisation varies between the
three instrument models), except for XLSSC 007 and 013, where the pn
model is displayed. The images are of size 'boxside' (see Table
\ref{tab_spatfit}).}
\label{sbfits}
\end{figure*}

Table \ref{tab_spatfit} lists the best--fit structural parameters for
each of the confirmed cluster sources.  Figure \ref{sbfits} shows flux
contours corresponding to the best--fitting surface brightness model
overplotted on the X--ray emission for each candidate.  The most
interesting outcome from these fits is the low value of the fitted
$\beta$ parameter for many of these systems, compared with the typical
value of $\beta=0.66$ determined for clusters (Arnaud \& Evrard
1999). The median value of $\beta=0.45$ found here agrees with the
value of 0.46 derived for a sample of low redshift groups by Heldson
\& Ponman (2000). Table \ref{tab_spatfit} also lists the value of the
normalised $\chi^2$ statistic computed from a comparison of the best
fitting model to the radially averaged surface brightness profile for
each source. The value of the $\chi^2$ statistic for each cluster
generally indicates that the computed model provides a statistically
acceptable fit. The two systems with computed $\chi^2$ values
significantly less than one (14 and 16) display some of the lowest
count levels in the sample and, in these cases, the radially averaged
counts may be better described by a Poissonian rather than Gaussian
noise distribution -- partially invalidating the application of a
$\chi^2$ merit function.

\begin{table}

\caption[]{Morphological X--ray parameters determined for confirmed
groups and clusters. Displayed errors are $1\sigma$ and, in the case
of cluster 19, no reliable error information could be
determined. Values of $\chi^2$ were computed over a sum of radial bins
for each cluster. These values are provided to indicate the overall
merit of each fit.}
\label{tab_spatfit}
\begin{tabular}{ccccc} \hline
Cluster & boxside & $r_0$ & $\beta$ & $\chi^2$ \\
& arcsec & arcsec & & (per d.o.f.) \\ \hline \hline
XLSSC 006   & 250    & $24.0_{-2.4}^{+3.1}$  & $0.58_{-.02}^{+.04}$  & 1.34\\
XLSSC 007   & 400    & $24.0_{-16.0}^{+14.0}$& $0.40_{-.06}^{+.07}$  & 1.15\\
XLSSC 008   & 250    & $9.6_{-6.0}^{+8.7}$   & $0.44_{-.06}^{+.09}$  & 1.26\\
XLSSC 009   & 250    & $44.0_{-19.0}^{+52.0}$& $0.67_{-.17}^{+.73}$  & 0.96\\
XLSSC 010   & 400    & $8.5_{-2.6}^{+3.3}$   & $0.44_{-.02}^{+.03}$  & 1.68\\
XLSSC 012   & 500    & $29.2_{-7.2}^{+10.0}$ & $0.54_{-.48}^{+.08}$  & 1.39\\
XLSSC 013   & 175    & $11.8_{-6.2}^{+14.9}$ & $0.67_{-.16}^{+.67}$  & 1.34\\
XLSSC 014   & 175    & $4.7_{-4.7}^{+39}$    & $0.40_{-.16}^{+.20}$  & 0.57\\
XLSSC 016   & 200    & $2.5_{-2.4}^{+3.5}$   & $0.45_{-.07}^{+.10}$  & 0.79\\
XLSSC 017   & 200    & $12.1_{-12.0}^{+7.6}$ & $0.55_{-.26}^{+.04}$  & 1.11\\
XLSSC 018   & 250    & $5.2_{-2.8}^{+3.4}$   & $0.42_{-.03}^{+.03}$  & 1.14\\
XLSSC 020   & 200    & $24.8_{-22.2}^{+8.1}$ & $0.40_{-.16}^{+.04}$  & 0.99\\ \hline
\end{tabular}
\end{table}

\subsection{Spectral properties}

\begin{figure}
\centering
\includegraphics[height=8cm]{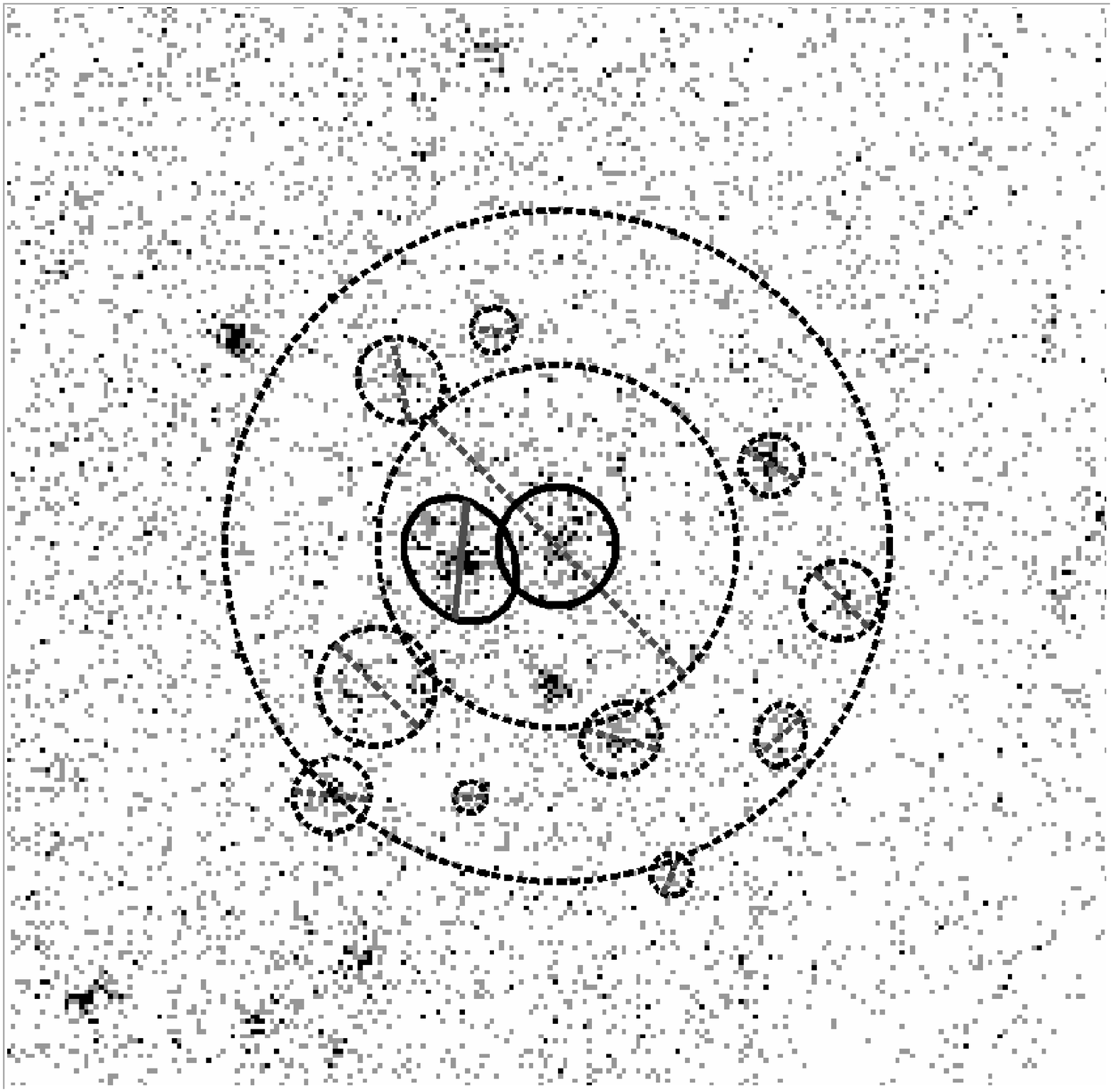}
\caption{Source and background extraction regions applied to
XLSSC 013. The greyscale shows the photon map generated by the stacked
pn $+$ MOS1 $+$ MOS2 image.The central solid circle represents the
source extraction region. Solid circles crossed with a diagonal line
indicate regions excluded from the source aperture. The two large
concentric dashed circles define the background annulus and dashed
circles crossed with diagonal lines indicate regions excluded from the
background aperture.}
\label{fig_specex}
\end{figure}

X--ray spectra for each cluster were extracted within a source circle
of radius ranging between 30\arcs\ and 90\arcs. The corresponding
background spectrum was extracted from a surrounding annulus.  Sources
adjacent to the cluster were flagged and removed from the spectral
analysis employing the source region file generated by the original
source extraction software. Source mapping used the stacked pn $+$
MOS1 $+$ MOS2 image of each cluster field in the [0.5-2]~keV band
only. The source extraction and background regions applied to
XLSSC 013 is shown in Figure \ref{fig_specex} as an example of the
procedure.

\begin{figure*}
\centering
\includegraphics[width=15.5cm]{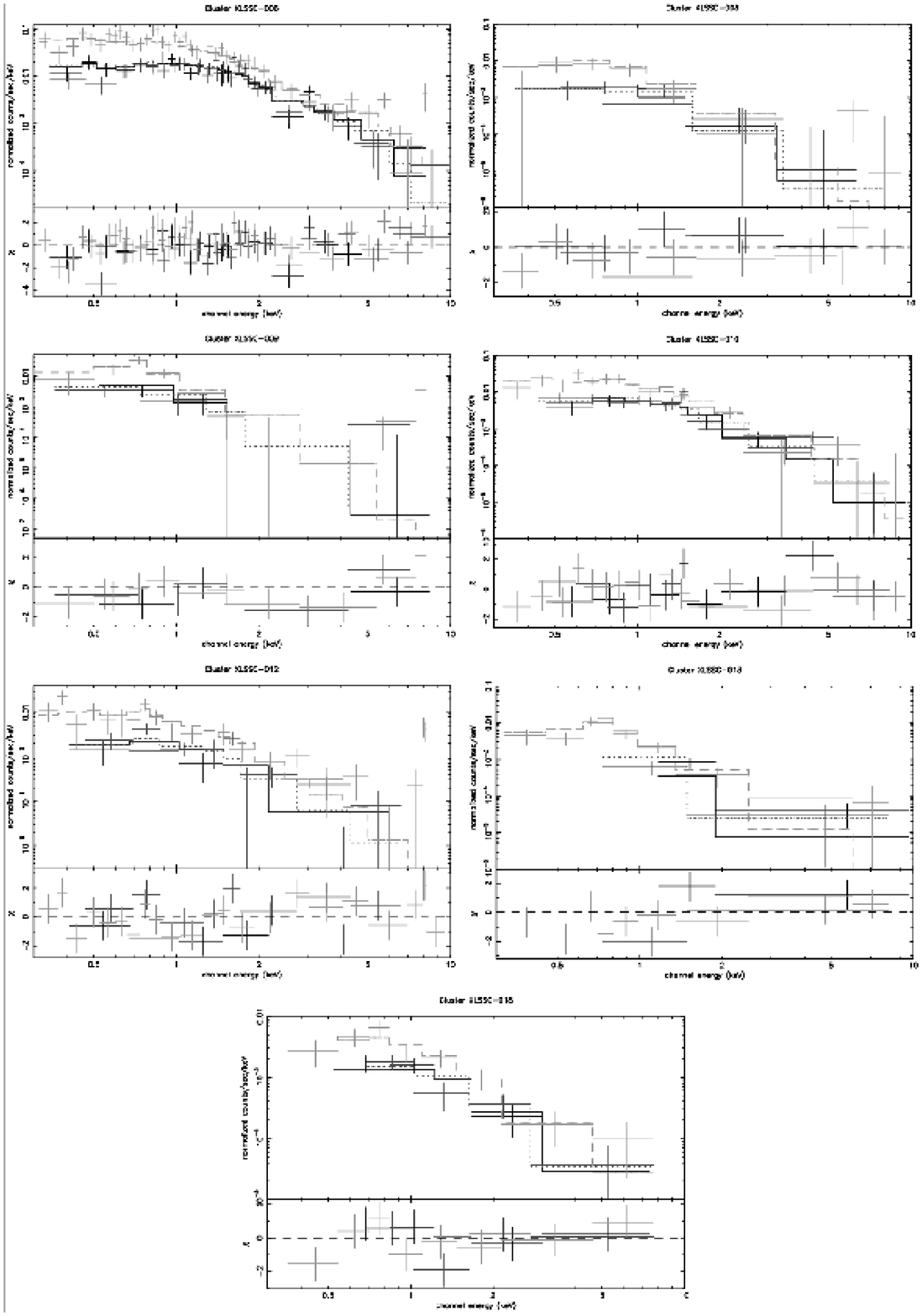}
\caption{Spectral data for groups and clusters described in Table
\ref{tab_specfit} for which a temperature was fitted. For each system
the upper panel depicts the spectral photon flux for the MOS1 (thick
black crosses), MOS2 (thin black crosses) and pn (grey crosses). The
data have been resampled to 20 photons per spectral bin for display
purposes only. The spectral model applicable to each detector is also
shown by the appropriately coloured solid line, i.e.  MOS1 (solid
black line), MOS2 (dotted black line) and pn (grey dashed line). The
lower panel shows the deviation of the data from the model for each
detector expressed in units of normalised $\chi^2$ per spectral bin.}
\label{xspecall}
\end{figure*}

Extracted spectral data corresponding to pn $+$ MOS1 $+$ MOS2
detectors were fitted simultaneously.  The fitting model consists of
an absorbed APEC hot plasma model (Smith et al. 2001) with a metal
abundance ratio set to Grevesse and Sauval (1999) values. The hydrogen
absorption is modeled using a {\it wabs} model with $N_H$ fixed at the
Galactic value, i.e.  $\rm N_H \sim 2.6\times
10^{20}$~cm$^{-2}$. Spectral data are resampled such that the
associated background spectrum displays 5 counts per bin and model
values are compared to the data by computing the corresponding value
of the C--statistic (see Appendix \ref{app} for a justification for
this approach). Model fitting is performed in two stages; first the
temperature and abundance are fixed ($T=0.5$~keV, $Z/Z_{\odot}=0.3$)
and only the count normalization is fitted.  Once a best--fitting
spectrum normalisation has been computed, the best--fitting
temperature is computed assuming a fixed metal abundance. The fitting
results are displayed in Table \ref{tab_specfit}. Table
\ref{tab_specfit} also shows the value of the normalised C--statistic
computed from a comparison of the spectral data for each cluster to
the quoted model. In each case a statistically reasonable agreement is
obtained. Spectral data for all clusters possessing a fitted
temperature in Table \ref{tab_specfit} are displayed in Figure
\ref{xspecall}.  In the cases of clusters 7, 14, 16, 17 and 20, no
spectral fit was possible and a temperature of 1.5 keV (typical of the
sample as a whole) was assumed for the purpose of calculating the flux
from the system.

X--ray luminosities for each cluster source have been determined
within a uniform physical scale derived from the cluster overdensity
radius. In the present study we employ the radius $r_{500}$, within
which the total mean density of the system is 500 times the critical
density of the universe at the redshift of the system. The value of
$r_{500}$ is computed using an isothermal $\beta$--model (Ettori 2000)
and employing the fitted gas temperature and $\beta$ value for each
system\footnote{We modify Equation A2 of Ettori (2000) to reflect the
varying redshift dependence of Hubble parameter in a matter plus
lambda cosmological model.}.

We have compared the value of $r_{500}$ derived from the above
isothermal model for each cluster to the corresponding values obtained
using the observed relationship between cluster mass and
temperature. Employing the data presented by Finguenov, Reiprich and
B{\"o}hringer (2001), based on systems ranging from $T_X=0.75-14$~keV,
we obtain the following relation
\begin{equation}
{
r_{500} = 0.391 \, T_X^{0.63} \, h_{70}(z)^{-1} \; \rm Mpc,
}
\end{equation}
where $h_{70}(z)$ describes the redshift evolution of the Hubble
parameter (scaled to 70 kms$^{-1}$ Mpc$^{-1}$) in the assumed
cosmological model. The value for $r_{500}$ for each cluster derived
using the above methods typically agree to within $\pm 10$\%\ with the
maximal difference being $\pm 20$\%. Based upon this comparison we
employ $r_{500}$ values based upon the isothermal model in the rest of
this paper.

In those cases where a successful spectral fit was obtained, the
derived bolometric source flux was extrapolated to $r_{500}$ using the
$\beta$--model determined in Section 5.1. For the four systems with
detected flux but with no successful spectral fit, the corresponding
best--fitting spatial model was employed to compute the bolometric
flux within $r_{500}$, assuming a $T=1.5$~keV emission
spectrum. Uncertainties on the resulting value of bolometric
luminosity are available only for those systems with spectral fits, in
which cases the uncertainty is simply derived from the error on the
fitted spectral model normalisation. Computation of luminosity
uncertainties employing this method does not include the effects of
the uncertain extrapolation of the surface brightness model to
$r_{500}$. The aperture correction factor for each cluster, $A$,
defined as the relative change in the integrated $\beta$--model
profile obtained by varying the integration limit from $r_{spec}$ to
$r_{500}$, is shown in Table \ref{tab_specfit}.

\begin{table*}
\centering
\caption[]{Spectral X--ray parameters determined for confirmed groups
and clusters. Values for exposure time, $t_{exp}$, and total counts
are summed over all three detectors. Where the letter ``F'' follows a
tabulated temperature (T) value, this indicates that the value was
fixed in the fitting procedure. The definition of the aperture
correction factor, $A$, is provided in the text. Displayed errors are
$1\sigma$.}
\label{tab_specfit}
\begin{tabular}{ccccccccc}\hline
Cluster & $t_{exp}$ & total & $r_{spec}$ & T  & C--stat & $r_{500}$ & $A$ & $L_{bol}(r_{500})$ \\
& seconds & counts & arcsec & keV & (per d.o.f.) & Mpc & & $\times 10^{43} \; \rm ergs \, s^{-1}$ \\ \hline \hline
XLSSC 006 & 17789 & 1943 & 82.5 & $4.80_{-0.84}^{+1.12}$  & 0.85 & 0.809 & 1.29 & $36.2 \pm 2.3$ \\
XLSSC 007 & 28094 & 138  & 90   & 1.5F                    & 1.10 & 0.284 & 0.65 & 1.1            \\				     
XLSSC 008 & 32358 & 94   & 60   & $1.25_{-0.38}^{+1.44}$  & 1.04 & 0.393 & 1.62 & $0.5 \pm 0.2$  \\
XLSSC 009 & 10709 & 112  & 90   & $0.91_{-0.17}^{+0.20}$  & 1.12 & 0.292 & 0.93 & $1.1 \pm 0.3$  \\
XLSSC 010 & 22635 & 505  & 67.5 & $2.40_{-0.53}^{+0.82}$  & 1.00 & 0.539 & 1.50 & $4.6 \pm 0.5$  \\ 
XLSSC 012 & 37726 & 635  & 60   & $2.00_{-0.51}^{+1.28}$  & 1.20 & 0.462 & 1.52 & $3.0 \pm 0.4$  \\ 
XLSSC 013 & 34383 & 133  & 35   & $1.03_{-0.25}^{+0.18}$  & 0.92 & 0.437 & 1.38 & $0.5 \pm 0.1$  \\ 
XLSSC 014 & 14801 & 286  & 50   & 1.5F                    & 1.26 & 0.404 & 1.59 & 0.4            \\ 				     
XLSSC 016 & 27202 & 25   & 30   & 1.5F                    & 0.99 & 0.432 & 1.76 & 0.4	           \\ 				     
XLSSC 017 & 25506 & 79   & 30   & 1.5F                    & 1.14 & 0.456 & 1.50 & 0.6	           \\ 				     
XLSSC 018 & 62573 & 295  & 45   & $2.66_{-0.91}^{+2.47}$  & 1.40 & 0.558 & 2.32 & $1.3 \pm 0.2$  \\ 
XLSSC 020 & 16770 & 61   & 37.5 & 1.5F                    & 1.09 & 0.305 & 1.01 & 2.0            \\ \hline 
\end{tabular}
\end{table*}

\section{The nature of XMM--LSS survey selected groups and clusters at  $z<0.6$}

\begin{figure}
\centering
\includegraphics[width=8cm]{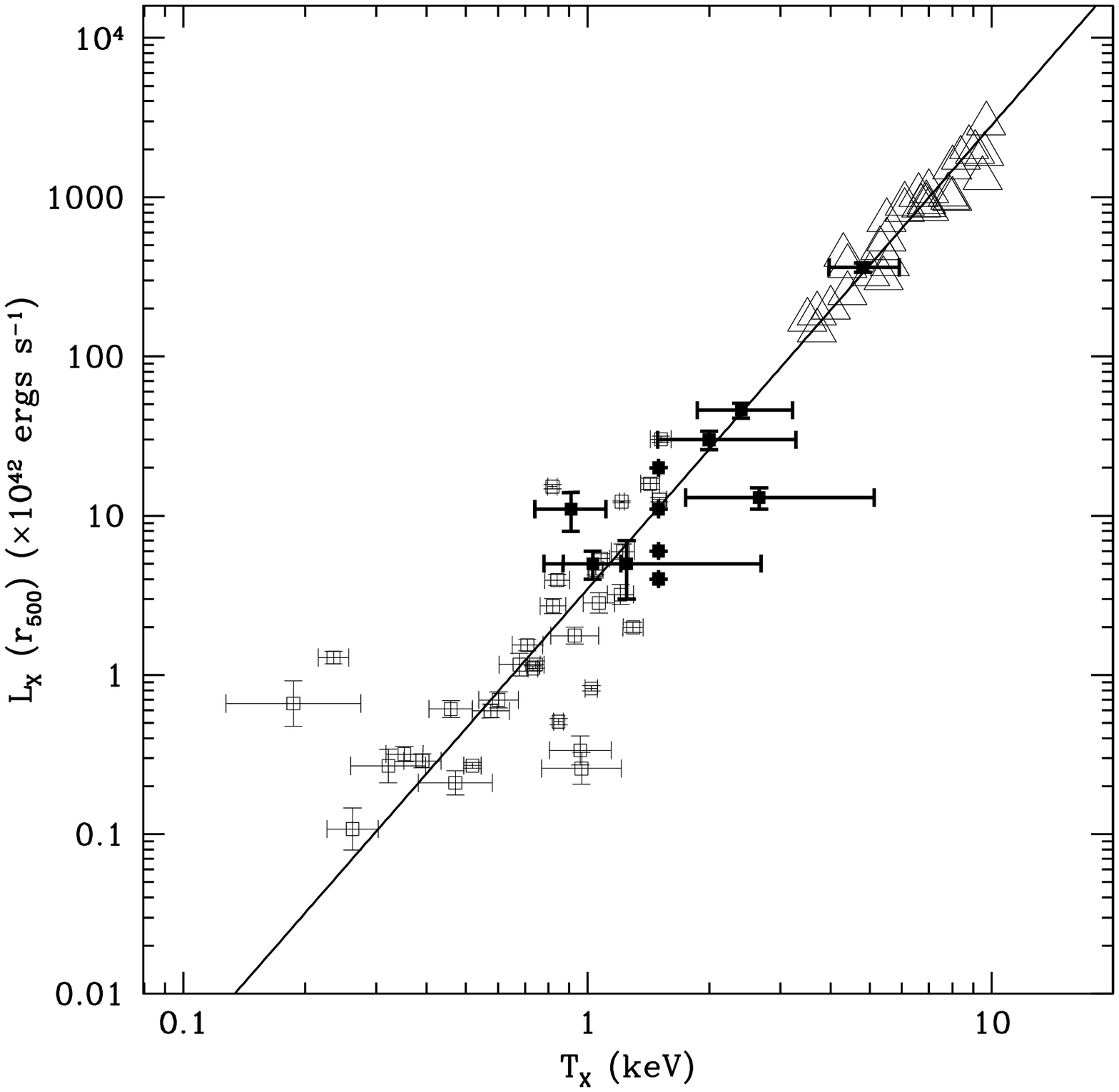}
\caption{Distribution of X--ray luminosity computed within a scale
radius $r_{500}$ and temperature for all XMM--LSS groups and clusters
currently identified at $z \le 0.6$ (solid squares). Also indicated
are values of X--ray luminosity and temperature determined for the low
redshift group sample of OP04 (open squares) and for the cluster
sample of Markevitch (1998) (open triangles). The solid line indicates
an orthogonal regression fit to the $L_X$ versus $T_X$ relation for
both the group and cluster sample incorporating a treatment of the
selection effects present in each sample (Heldson and Ponman 2005) --
see text for details.}
\label{lx_tx}
\end{figure}

In this section we compare the properties of the XMM--LSS groups and
clusters at $z < 0.6$ with X--ray group and cluster samples in the
literature.  Figure \ref{lx_tx} compares the $L_X$ versus $T_X$
distribution of XMM LSS clusters confirmed at $z<0.6$ to both the
distribution formed by the Group Evolution Multi--wavelength Study
(GEMS) sample of local ($z<0.03$) X--ray emitting galaxy groups
(Osmond and Ponman 2004; hereafter OP04) and the sample of Markevitch
(1998; hereafter M98) containing clusters at $z<0.09$ with ASCA
temperatures and ROSAT luminosities\footnote{Note that, although
luminosities in M98 are quoted within 1 $h_{100}^{-1}$ Mpc apertures
and not $r_{500}$ as used in this paper, the appropriate correction
factors are typically considerably smaller than the 5\%\ calibration
uncertainty associated with the luminosities themselves.}.  Figure
\ref{lx_tx} indicates that XMM--LSS clusters occupy a region of the
$L_X$ versus $T_X$ plane ranging from cool ($T_X > 0.9$ keV), low
luminosity ($L_X (r_{500}) > 4 \times 10^{42}$ ergs s$^{-1}$) X--ray
groups to moderate temperature ($T_X = 5$ keV), moderate luminosity
($L_X (r_{500}) = 4 \times 10^{44}$ ergs s$^{-1}$) clusters. Though
the sample of X--ray systems presented in this paper is not
statistically complete, it is representative of the properties of the
complete flux--limited sample currently under construction. Due to the
steeply rising nature of the XLF toward faint X--ray systems, it is
anticipated that the larger, statistically complete sample of X--ray
structures identified by XMM--LSS at $z<0.6$ will be dominated by such
galaxy group and low mass cluster systems.

\begin{figure}
\centering
\includegraphics[width=8cm]{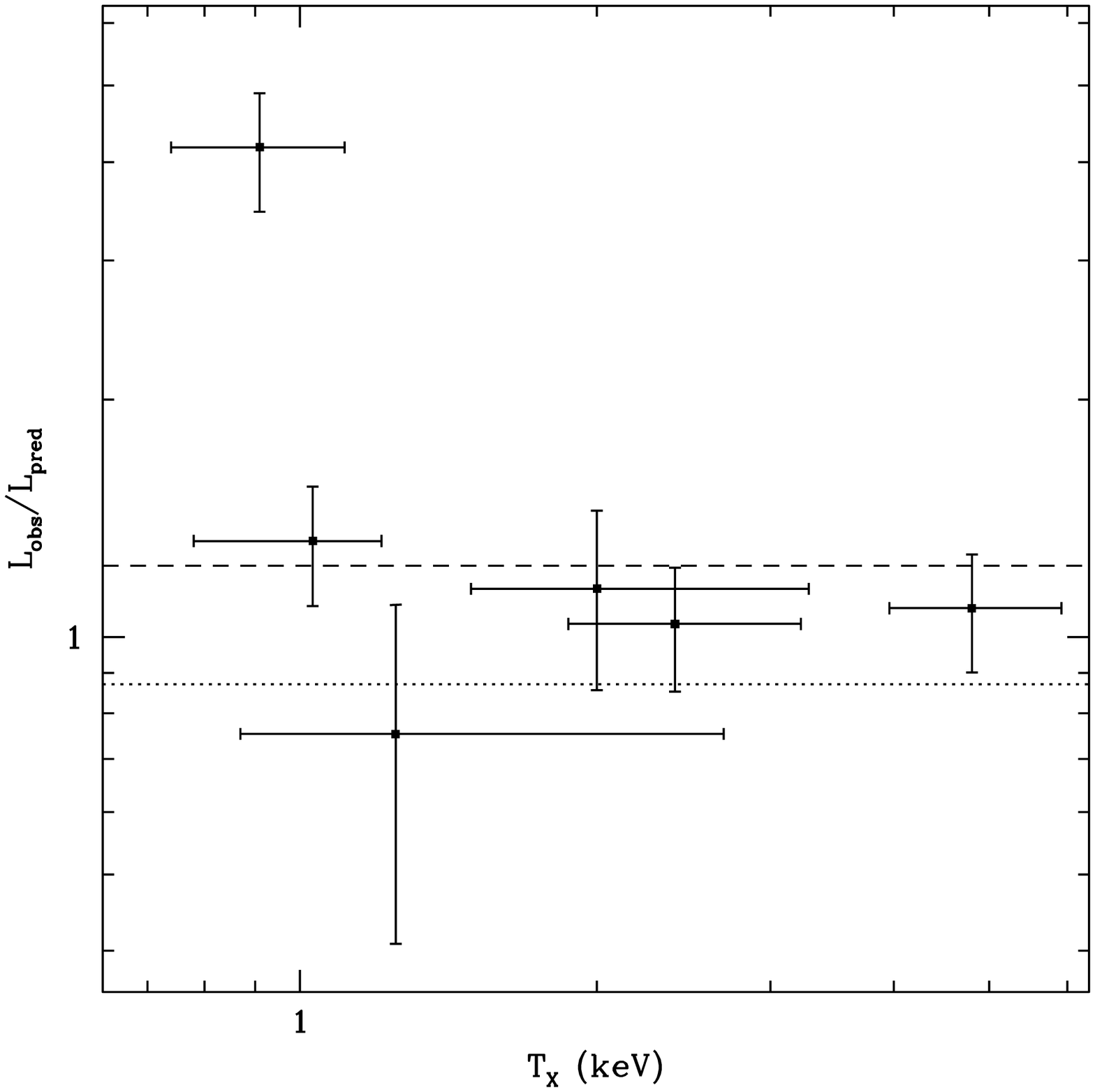}
\caption{Enhancement factor, $F = L_{obs}/L_{pred}$, computed for six
XMM--LSS groups and clusters located at $z \le 0.6$ plotted versus the
X--ray temperature of each system (see text for additional
details). Horizontal lines indicate expected values of $F$: the short
dashed line indicates the value $F=1.23$ expected from self--similar
considerations. The dotted line indicates the value of $F$
expected at $z=0.4$ based upon Ettori et al. (2004).}
\label{enhance_vs_tx}
\end{figure}

It can be seen from Figure \ref{lx_tx}, that our XMM--LSS groups and
clusters appear to be in good agreement with a linear fit to the $L_X$
versus $T_X$ distribution of lower redshift group and cluster samples.
The local fit to the $L_X$ versus $T_X$ distribution takes the form
$\log L_X = 2.91 ~\log T_X + 42.54$ and is derived from an orthogonal
regression fit to the combined OP04 plus M98 samples incorporating a
treatment of the selection effects present in each sample (Heldson and
Ponman 2005). The comparison of XMM--LSS groups and clusters to this
local relationship may be quantified (Figure \ref{enhance_vs_tx}) by
calculating a luminosity enhancement factor, $F=L_{obs}/L_{pred}$,
where $L_{obs}$ is the observed cluster X--ray luminosity within a
radius, $r_{500}$, and $L_{pred}$ is the luminosity expected applying
the fitted $L_X$ versus $T_X$ relation computed for the local fit and
the XMM--LSS measured temperature. Neglecting the 5 systems assigned a
fixed temperature (XLSSC 007, 014, 016, 017 and 020 -- for which the
temperature uncertainty is unknown) , the median enhancement factor of
the remaining 6 systems is $F = 1.09$. For comparison, the expected
enhancement in $L_X$ due to self-similar evolution, scaling to
$r_{500}$, is a factor of 1.23 at the typical redshift ($z=0.4$) of
our sample. Therefore, given the observed spread in the enhancement
values, the observed negative deviation from the self--similar
expectation is not large.

Given the modest size and the statistically incomplete nature of our
current sample, we regard these results as in need of confirmation.
Ettori et al. (2004) also report evolution weaker than the
self--similar expectation from a sample of 28 clusters at $z>0.4$ with
gas temperatures $3~{\rm keV} < T < 11~{\rm keV}$. The combined effect
of self--similar scaling with the negative evolution reported by
Ettori et al. (2004) would result in an enhancement factor
$F=0.86$\footnote{This enhancement factor assumes self--similar
evolution and an additional factor $(1+z)^{B_z}$ where $B_z = -1.04$,
following the nomenclature of Ettori et al. (2004).} at a $z=0.4$ (see
Figure \ref{enhance_vs_tx}). Though the overlap of the Ettori et
al. (2004) sample and the systems contained in the present work is
limited, the X--ray luminosities appear to describe similar trends.

\begin{figure}
\centering
\includegraphics[width=8cm]{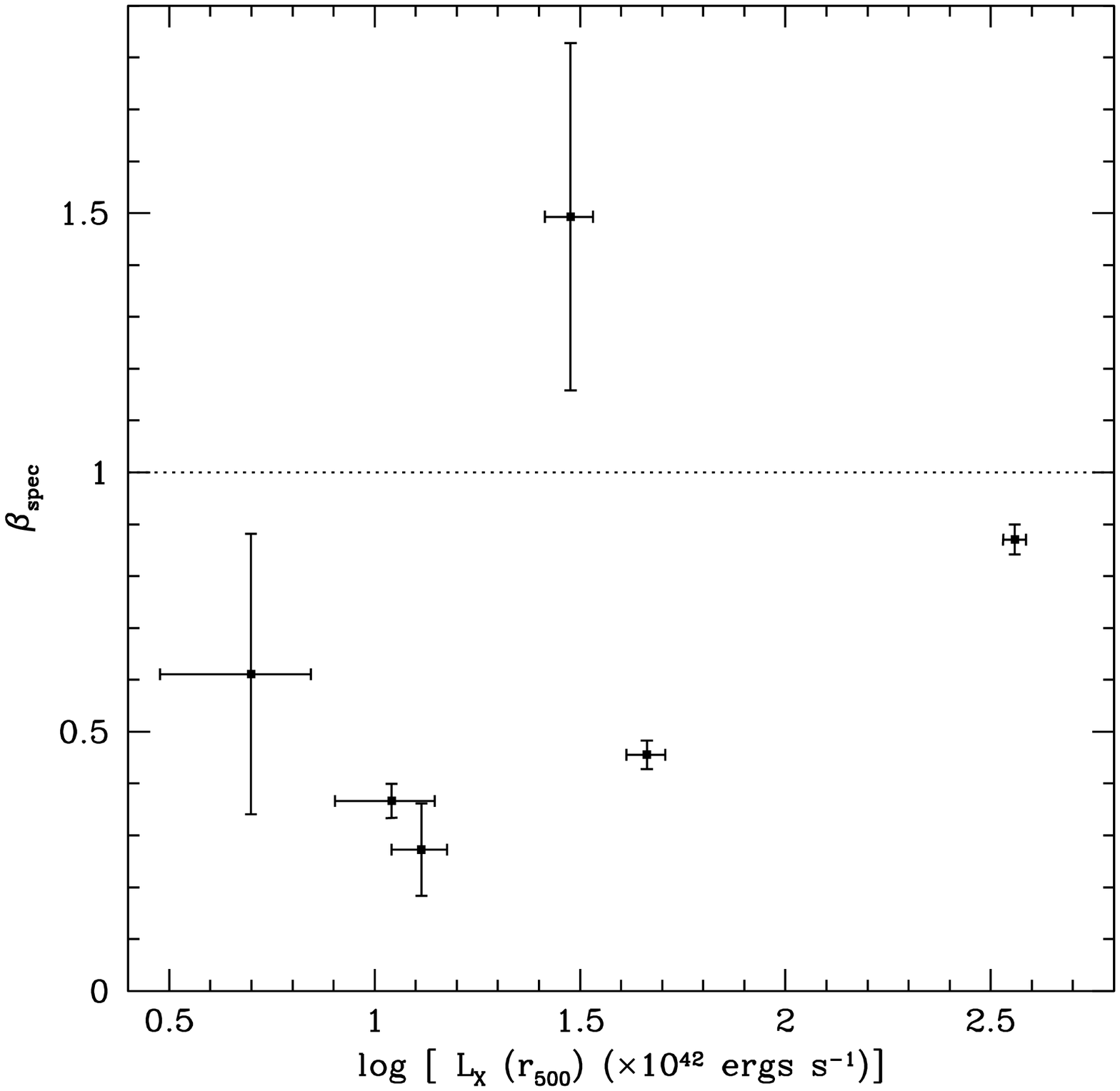}
\caption{Values of $\beta_{spec}$ computed for eight XMM--LSS groups
and clusters at $z \le 0.6$ (see text for details) plotted versus the
X--ray luminosity for each system. The horizontal dashed line
indicates the value $\beta_{spec} = 1$.}
\label{bspec_vs_lx}
\end{figure}

The relationship between the specific energy contained within the
cluster galaxy motions, compared to the X--ray emitting gas, is described
via the $\beta_{spec}$ parameter
\begin{equation}
{ 
\beta_{spec} \equiv \frac{\sigma_v^2}{kT_X/\mu m_p}
}
\end{equation}
where $\sigma_v$ is the line--of--sight galaxy velocity distribution,
$k$ is the Boltzmann constant, $T_X$ is the X--ray gas temperature and
$\mu m_p$ is the mean particle mass within the gas (Bahcall \&\ Lubin
1994). Figure \ref{bspec_vs_lx} displays the value of $\beta_{spec}$
computed for the XMM--LSS $z<0.6$ cluster sample as a function of
computed X--ray luminosity extrapolated to a radius $r_{500}$
(i.e. the consistent measure adopted in this paper).  A number of
systems have been excluded from Figure \ref{bspec_vs_lx}: in addition
to the systems excluded from the computation of the luminosity
enhancement factor above, XLSSC 013 does not possess well defined
galaxy velocity dispersion thus preventing computation of
$\beta_{spec}$. Figure \ref{bspec_vs_lx} indicates typical values
$\beta_{spec}<1$ for the XMM--LSS $z<0.6$ sample (with the exclusion
of the above mentioned systems), the median value of $\beta_{spec}$
for this restricted sampled is $\langle \beta_{spec}=0.61
\rangle$. This value may be compared to values of $\beta_{spec}
\approx 1$ reported by OP04 for luminous X--ray groups (i.e. $L_X
(r_{500}) > 10^{42}$ ergs s$^{-1}$).

A clear concern when interpreting the trend of low $\beta_{spec}$
values is the extent to which the cluster galaxy velocity dispersion
estimates may be biased toward lower values. Potential uncertainties
associated with the velocity dispersion values presented in this paper
are discussed in Section \ref{sec_clsprop}. However, in the
overwhelming majority of clusters observed in detail, the integrated
velocity dispersion profile of galaxy clusters is a {\it decreasing}
function of projected radius from the cluster centre (Girardi et
al. 1996; Borgani et al. 1999). If the integrated velocity dispersion
profiles of the XMM--LSS clusters presented in this paper display
similar behaviour to hotter clusters, then the expectation arising
from computation of the cluster velocity at some fraction of the
convergent radius is that the velocity dispersion will be
overestimated and will result in values of $\beta_{spec}$ biased to
higher values. The extent of any such bias is difficult to quantify in
the current data set. However, the implication is that the value of
$\beta_{spec}$ displayed in Figure \ref{bspec_vs_lx} would not
increase with the addition of velocity dispersion measurements
extending to larger radii.

The low values of $\beta_{spec}$ apparent in Figure \ref{bspec_vs_lx},
are similar to those seen by OP04 in lower luminosity groups ($L_X <
10^{42}$ ergs s$^{-1}$) at low redshift. The origin of these low
values of $\beta_{spec}$ is far from clear, but OP04 argue that it
appears to result primarily from a reduction in $\sigma_v$, rather
than an enhancement in $T_X$. Whatever the cause, our results provide
tentative evidence that these effects are operating in hotter and more
X--ray luminous systems at higher redshift. We are currently in the
process of conducting magnitude limited spectroscopy of a sample of
$T_X \sim 1$~keV systems at $z=0.3$ in order to provide a more robust
picture of the dynamics of low temperature X--ray systems.

\section{Conclusions}

We have presented twelve newly identified X--ray selected groups and
clusters as part of the XMM Large--scale Structure Survey. The
procedures employed to detect and classify sources in X--rays, and to
subsequently confirm each source via optical imaging and spectroscopic
observations have been described in detail.

We have emphasized throughout this paper that the current sample of
clusters is not complete in any statistical sense. The presentation of
a larger, complete sample of X--ray clusters located at $z<0.6$ will
form part of a future publication.  However, the current sample of
X-ray clusters at $z<0.6$ presents a number of interesting features:
most importantly, the sample is dominated by low X--ray temperature
systems located at redshifts much greater than that presented by
previous X--ray studies. Such systems are predicted to display the
effects of pre--heating or additional energy input into the ICM to a
greater extent than hotter, more massive systems. The identification
of such low--temperature systems at look--back times up to 5.7 Gyr
provides an important baseline over which to study the extent to which
such systems evolve.

We find tentative evidence that these high redshift groups are more
luminous than local systems, at a given temperature, in agreement with
recent work on richer clusters.  However, our results suggest that
group luminosities may be evolving less rapidly than higher
temperature clusters when compared to self-similar models.  If this is
confirmed to be the case, then the steepening of the $L_X$-$T_X$
relation at low temperatures reported in local samples, may continue
at higher redshift.  We also find preliminary indications that the
poorly understood tendency for the specific energy in the gas to
exceed that in the galaxies in poor groups, extends to systems with
higher values of $L_X$ and $T_X$ at $z\sim0.4$. The completion of a
larger and statistically complete sample of intermediate redshift
groups from the XMM--LSS survey, should allow these results to be
placed on a firm statistical footing in the near future.

\section*{Acknowledgements}

The authors gratefully acknowledge Steve Heldson for his assistance
with comparisons with the low redshift X--ray group and cluster
samples. The authors additionally thank Jean Ballet and Keith Arnaud
for useful discussions on the statistical treatment inside {\tt Xspec}
and Jean-Luc Sauvageot for technical discussions regarding XMM
calibration for fitting purposes.

\appendix

\section[]{Fitting simulated cluster spectra}
\label{app}

The current study extends X--ray spectral observations of distant
galaxy groups and clusters to low integrated signal levels ($\sim100$
photons above the background).It is therefore prudent to assess the
reliability of temperature measures and associated uncertainties
computed via model fits to such faint spectra by repeating the fitting
procedure for a grid of simulated spectra created to reproduce the
properties of the observed sample.

The source model used to simulate group and cluster spectra employs an
APEC of an optically thin plasma (Smith et al. 2001). This model
depends upon four parameters: temperature, metal abundance, redshift
and a normalisation representative of the emission integral. To
simulate galaxy groups and clusters the abundance is set to $Z/Z_\odot
= 0.3$ with solar abundance ratios set to Grevesse and Sauval (1999)
values. The redshift of the simulated source is set to $z=0.3$,
typical of the sources presented in this paper. Photoelectric
absorption described by a {\it wabs} model within {\tt Xspec} using
Morrison and McCammon (1983) cross--sections was applied to the source
model with the neutral hydrogen column density fixed to $\rm N_H =
2.6\times 10^{20}~ cm^{-2}$ -- the mean value for our sample according
to the H{\sevensize I} distribution map of Dickey and Lockman
(1990). The instrumental response was modeled using the redistribution
matrix and ancillary response files from observations of XLSSC 006.

A model pn $+$ MOS1 $+$ MOS2 spectra was created and a conversion
factor applied to generate spectra of the required integrated count
level over the spectral interval [0.3--10]~keV in a 10,000 second
exposure. Each simulated spectrum is generated from this model using
Poissonian considerations.  Each source spectrum is accompanied by a
background spectrum created from a Poisson realisation of a background
model of normalisation and shape consistent with observed cluster
backgrounds. Spectra were simulated according to this procedure for
temperatures equal to $1,2,3$ and 5~keV. At each temperature, 50
spectra were simulated at each point of a grid of integrated count
levels $100, 200, \dots, 1000$.

\begin{figure*}
\centering
\includegraphics[width=17cm]{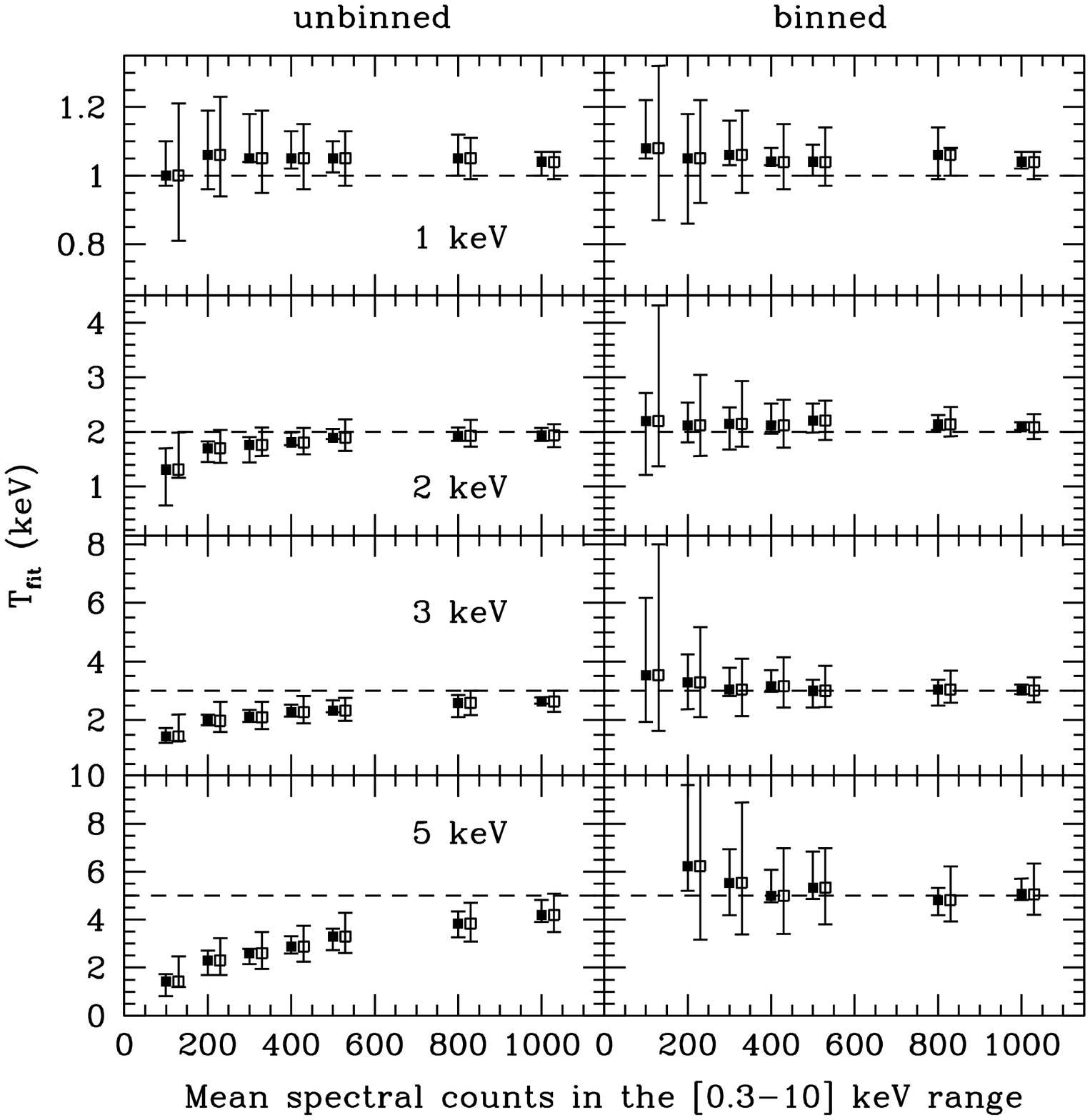}
\caption{A comparison of {\tt Xspec} computed temperatures for
simulated group and cluster spectra employing the C--statistic with
two different resampling approaches. Left panels indicate the results
for unbinned spectra. Right panels indicate the results for binned
spectra such that the background spectrum displays a minimum of 5
counts per spectral bin. Panels in each row correspond to spectral
models with the indicated input temperature (also shown by the
horizontal dashed line). In each panel, data points represent the mean
{\tt Xspec} computed temperature returned from the set of simulated
spectra as a function of total input counts. Filled squares plus error
bars indicate the mean computed temperature and the distribution of
temperatures accounting for 68\%\ of the sample. Open squares plus
error bars indicate the mean computed temperature and the median
$1~\sigma$ uncertainty returned by {\tt Xspec} (open squares are
shifted to the right by 30 counts with respect to the filled squares
for clarity).}
\label{tempsim}
\end{figure*}

Spectral fitting follows the same approach as applied to observed
data, i.e. temperature and spectum normalisation are permitted to vary
while the abundance is fixed. Data from pn $+$ MOS1 $+$ MOS2 are are
combined within {\tt Xspec} with the response files from XLSSC 006
and the energy range [0.3--10]~keV is conserved. The spectral energy
range corresponding to [7.5--8.5]~keV measured by the pn detector is
ignored as it contains strong instrumental line emission that is not
well corrected by our data modeling process. The best--fitting model
is then determind by minimising a modified C--statistic and the
$1~\sigma$ uncertainty about the best fit model is computed. While
the C--statistic is intended to work efficiently on unbinned data a
comparison of the fitted temperature to the input value indicates a
tendency to underestimate the temperature of spectra of input
temperatures $<5$~keV displaying count values $<1000$ counts using
this procedure. This bias is indicated in Figure \ref{tempsim} and
appears to arise from the fact that the statistic used in {\tt Xspec}
represents a modified C--statistic that accounts for statistical
fluctuations in the background estimation. Our understanding of the
problem is that this modified statistic fails at estimating model
parameters when there is a significant number of background bins
containing zero photons.

Resampling the data to prevent the occurence of spectral bins
containing zero counts minimises this negative temperature bias. The
resampling factor is determined by requiring that the background
spectrum associated with each source display a specified minimum count
level per spectral bin. Determining the resampling factor from the
background spectrum represents a sensible approach as the background
counts are more numerous and therefore minimise the loss of spectral
information in the source spectrum. Although a small positive bias is
introduced to the fitted temperatures of very low count level spectra
($<300$ counts) when the data are resampled, the amplitude of this
bias is less than 10\%\ when the data are resampled to contain 5
counts per spectral bin (Figure \ref{tempsim}).  Applying a larger
resampling factor increases the positive temperature bias -- which can
be understood in terms of the spectral smoothing that the resampling
procedure represents.

We therefore resample the observed data to generate a background
spectrum containing 5 counts per spectral bin. This approach generates
fitted temperatures that agree with the input temperature to
$<10$\%. In addition, comparison of the distribution of fitted
temperatures at any given combination of temperature and count level
to the temperature uncertainty returned by {\tt Xspec} indicates that
the {\tt Xspec} quoted errors on fitted temperatures overestimate the
distribution of fitted temperatures by a factor typically less than
2. Due to the various assumptions that enter the simulation procedure
it therefore seems reaonable to provide {\tt Xspec} quoted temperature
uncertainties as a conservative error estimate.

\bsp

\label{lastpage}

\end{document}